\documentclass{ar-1col-S2O}
\usepackage[numbers]{natbib}
\usepackage{url}
\usepackage{mystyle}
\jname{Xxxx. Xxx. Xxx. Xxx.}
\jvol{AA}
\jyear{YYYY}
\doi{10.1146/((please add article doi))}

\begin{document}

% Page header
\markboth{Esterlis and Schmalian}{QC Eliashberg Theory}

% Title
\title{Quantum Critical Eliashberg Theory}

%Authors, affiliations address.
\author{Ilya Esterlis$^1$ and J{\"org} Schmalian$^2$ 
\affil{$^1$Department of Physics, University of Wisconsin-Madison, Madison, Wisconsin 53706, USA; email: esterlis@wisc.edu}
\affil{$^2$Institute for Theory of Condensed Matter and Institute for Quantum Materials and Technologies, Karlsruhe Institute of Technology, Karlsruhe 76131, Germany; email: joerg.schmalian@kit.edu}}

%Abstract
\begin{abstract}
Quantum criticality plays a central role in understanding non-Fermi liquid behavior and unconventional superconductivity in strongly correlated systems. In this review, we explore the quantum critical Eliashberg theory, which extends conventional Eliashberg approaches to non-Fermi liquid regimes governed by critical fluctuations. We discuss the theoretical foundations and recent developments in the field, focusing on the interplay between electronic interactions and bosonic modes near quantum phase transitions as described in the Yukawa-coupled version of the Sachdev-Ye-Kitaev model.  Special emphasis is placed on the breakdown of quasiparticle coherence, anomalous scaling behaviour, Cooper pairing without quasiparticles, and emergent universality in different physical settings.  Starting from a zero-dimensional “quantum-dot” model, we discuss the generalization to higher spatial dimensions and demonstrate the connection between quantum-critical Eliashberg theory and holographic superconductivity.  Our analysis provides a perspective on how quantum criticality shapes the dynamics of strongly correlated metals and superconductors.
\end{abstract}

%Keywords, etc.
\maketitle

%Table of Contents
\tableofcontents

% Heading 1
\section{INTRODUCTION}

Despite the presence of a large number of gapless degrees of freedom, the metallic state is remarkably stable. In the language of Fermi-liquid theory \cite{landau1957,agd,baym2008landau,polchinski1992,shankar1994}, this stability is a consequence of kinematics of excitations near the Fermi surface and the Pauli principle. Nevertheless, at sufficiently low temperatures, the ultimate fate of most metallic systems is to become superconducting \cite{bcsshort,bcslong,kohn1965}. This outcome is a consequence of the one generic instability of the Fermi-liquid metal: the Cooper instability \cite{Cooper1956,agd,polchinski1992,shankar1994}.

While the Bardeen Cooper Schrieffer (BCS) superconductivity that emerges from a conventional Fermi-liquid is well understood, there is ample experimental evidence that some of the most interesting superconductors -- such as high-$T_c$ cuprates \cite{legros2019}, iron-based superconductors \cite{hayes2016, jiang2023}, heavy fermion compounds \cite{nguyen2021}, and others \cite{lee2023, jaoui2022} -- emerge from ``strange metal" states, whose phenomenology does not conform to the expectations of the conventional Fermi-liquid theory of metals. Among possible scenarios for destabilizing the Fermi-liquid state, the coupling of gapless metallic degrees of freedom to soft order parameter fluctuations near a \textit{quantum critical point} (QCP) is an especially interesting possibility that may be relevant to a variety of these materials. Indeed, in many cases the superconducting $T_c$ is maximal just above the QCP. Near such ``metallic quantum critical points",  the singular interaction between electrons mediated by the gapless order parameter fluctuations leads to significant quasi-particle damping, rendering the Fermi-liquid picture invalid.  The question then arises: at low temperatures, does such a ``non-Fermi liquid" (NFL) state \cite{lee2018} undergo a superconducting transition and, if so, what is the nature of the superconducting state?

The interplay between metallic quantum criticality, non-Fermi liquids, and superconductivity has been studied for many years, with significant progress having been made across several decades. The instability of the Fermi liquid near a density-wave QCP (for example, a charge or spin-density wave transition) has been investigated \cite{millis1993,altshuler1995, castellani1995,abanov2000, abanov2003,pankov2004, chubukov2005, lohneysen2007, metlitski2010b, efetov2013,abrahams2014,meier2014,varma2015,schlief2017,lunts2017}, as well as near QCPs associated with uniform order (such as nematic or ferromagnetic transitions) \cite{oganesyan2001, metzner2003,pankov2004, lawler2006, rech2006, aji2007,zacharias2009, metlitski2010, maslov2010, dalidovich2013,fitzpatrick2014}. More generally, critical Fermi surfaces may also arise due to coupling to gauge fields, as in the half-filled Landau level or spinon Fermi surface \cite{lee1989gauge, halperin1993,polchinski1994low,nayak1994,chakravarty1995,bonesteel1996, lee2009,mross2010,holder2015}. The absence of a Cooper instability out of a quantum-critical NFL with instantaneous attraction was demonstrated early on \cite{balatsky1993, sudbo1995, yin1996}. The situation is remedied, however, when one takes into account that the pairing interaction itself becomes singular, thereby compensating the weakened pairing ability of the NFL metal. 

The systematic study of such ``quantum critical superconductors'' was initiated in a series of seminal papers \cite{bonesteel1996, son1999, abanov2001}, with the analysis being subsequently taken up in earnest across the various quantum critical systems mentioned above \cite{abanov1999, abanov2001, abanov2001b, roussev2001, abanov2003, abanov2004, chubukov2005, she2009, moon2010, levchenko2013,wang2013,wang2015,varma2016,khodas2020,lederer2015, metlitski2015,fitzpatrick2015,raghu2015,mandal2016,she2009, she2011, wang2017_eliash, wang2018_thermal,wang2016, wu2019, abanov2020-I,abanov2020-II,wu2020-III,wu2021-IV,wu2021-V,zhang2021-VI,zhang2023,nosov2023,abanov2025}.
In addition to analytic progress, significant insights into the problem of quantum critical metals and superconductivity have also been gained from numerically exact determinant quantum Monte Carlo (DQMC) simulations on a variety of minus-sign-problem free models \cite{berg2012, schattner2016, schattner2016_ising, dumitrescu2016,gerlach2017, lederer2017, li2017, wang2017, xu2017, wang2018, berg2019, klein2020, xu2020, lunts2023,patel2024_dqmc}. 

In the last several years, a new approach to the problem of quantum critical superconductivity has emerged \cite{esterlis2019,wang2020,hauck2020,wang2020b,classen2021,esterlis2021,guo2022,patel2023,valentinis2023_lett,valentinis2023,guo2024_cyclotron,li2024,pan2021,wang2021_dqmc,guo2024,sutradhar2024}, inspired by the ``Sachdev-Ye-Kitaev'' (SYK) model  \cite{sachdev1993,georges2000,sachdev2010,kitaev2015,kitaev2015b} and its generalizations \cite{sachdev2015,maldacena2016,polchinski2016,fu2017,bi2017,song2017,chowdhury2018,chowdhury2022,sachdev2024}. The approach is based on a large-$N$ limit, in which electrons and bosons (which can represent phonons, order parameter fluctuations in the case of a QCP, or gauge field fluctuations), interacting via a Yukawa-type coupling, are dressed with ``flavor indices''. When the Yukawa couplings between the electrons and bosons are taken to be random in the flavor space, taking the limit of a large number of flavors leads to a closed-form solution for the dynamics of the model. This class of models have come to be known as ``Yukawa-SYK" (YSYK) models.  Remarkably, the equations that emerge are a version of the familiar Migdal-Eliashberg equations of electron-phonon superconductivity \cite{migdal1958,eliashberg1960,marsiglio2020}, but now obtained as the exact solution in an appropriate large-$N$ limit. The equations can be solved with standard methods to obtain the properties of the quantum critical system in both the normal and superconducting states. Moreover, the approach appears to be free of the pathologies associated with expansions based on a large-$N$ number of fermion flavors \cite{lee2009} and, unlike certain matrix large-$N$ approaches \cite{fitzpatrick2014}, incorporates important effects such as Landau damping of the boson. 

The YSYK approach has provided a controlled framework within which to analyze quantum-critical metals and superconductors, unifying different analytical approaches to the problem and leading to new insights into quantum critical superconductivity and non-Fermi liquids more broadly. Additionally, the  intimate connection between SYK models and gravitational theories in one higher dimension through the holographic principle~\cite{Maldacena1998,Witten1998,Gubser1998} has been used to construct an explicit mapping between the superconducting YSYK model and dual gravity theories~\cite{Inkof2021,2209.00474}. These gravity theories can be understood as a generalization of the Landau-Ginzburg theory of superconductivity extended into the quantum critical regime. Most recently, YSYK models with certain types of spatial disorder have been demonstrated to reproduce much of the salient transport phenomenology of superconductors emerging from candidate non-Fermi liquid states~\cite{esterlis2021,guo2022,patel2023,li2024}. 

This review article summarizes the YSYK approach to quantum-critical superconductivity, highlighting key results and describing general lessons that have been learned. In Section \ref{sec:quantum_dot} of this article, we start by describing the (0+1)-dimensional YSYK quantum dot model, including both its normal state and superconducting properties. In Section \ref{sec:higher_dim}, we discuss the generalization to higher spatial dimensions and the connection between those theories and documented phenomenology of strange-metal superconductors. Section \ref{sec:holography} reviews the connection between the YSYK models and holographic superconductivity, highlighting key steps in the explicit mapping. Our conclusions and outlook are provided in Section \ref{sec:conclusions}.  

Since the introduction of the YSYK model, a number of extensions and generalizations to other contexts have also been made. An incomplete list includes: connections to thermalization and hydrodynamics \cite{hosseinabadi2023_thermalizationNFL}, anomalous critical behavior in Dirac fluids~\cite{kim2021dirac}, non-equilibrium superconductivity \cite{grunwald2024_light-inducedSC}, inclusion of dissipation \cite{cichutek2024_dissipativeSYK}, non-Fermi liquids coupled to dynamical two-level systems modeling metallic glasses \cite{bashan2024_2-level-sys,tulipman2024solvable}, quantum chaos \cite{Tikhanovskaya2022_maximalchaos,davis2023_chaos}, pair-density wave order \cite{wu2023_pdw}, Andreev scattering at superconducting puddles \cite{bashan2025extended}, and studies of collective modes \cite{wang2023_density,wang2024_landau-damp}. We also highlight other SYK-inspired models of superconductivity in Refs.~\cite{patel2018, gnezdilov2019, chowdhury2020, wang2020_kuramoto,lantagne-hurtubise2021_spinfulSYK, choi2022_SC, chudnovskiy2022_SC-insulator-transition, gnezdilov2022_4e, li2023_SYKSC}.

%Heading 1
\section{Quantum dot}
\label{sec:quantum_dot}
The original (0+1)-dimensional model adopted in Refs.~\cite{esterlis2019,wang2020,hauck2020} aimed to combine the well-known tendency of electron-phonon (or more general boson) systems toward superconductivity together with the controlled large-$N$ SYK approach to generating non-Fermi liquid (singular self-energy) effects for electrons \cite{chowdhury2022,sachdev2024}. While a (0+1)-dimensional  ``quantum dot'' system does not possess a Fermi surface, it is nevertheless a simple starting point in which to study the interplay of superconductivity and singular interaction effects. Moreover, it exhibits the highly unusual feature of ``self-tuned criticality," wherein the system naturally becomes critical at low energies independent of the bare parameters. In addition, the quantum dot theory displays a pairing instability that also occurs in higher-dimensional problems whenever the frequency dependence of the electronic self-energy dominates over its momentum dependence.  We will thus start by describing the quantum dot model and its salient features. We will then discuss generalizations to higher dimensions \cite{esterlis2021,guo2022,patel2023,valentinis2023_lett,valentinis2023,guo2024_cyclotron,li2024}. When we discuss the relation to holography we will return to the quantum dot problem.
In all cases the degrees of freedom will be fermions $\psi_{i\sigma}$ where $\sigma$ is a physical spin index and $i=1,\ldots,N $ is a ``flavor index" that will be used to take a large-$N$ limit, as well as bosons $\phi_j$, with the flavor index $j=1,\ldots,M$. The number of boson and fermion flavors may be different in general, and the ratio $M/N$, which is kept fixed as $N\rightarrow \infty$, serves as an additional parameter in the model.

\subsection{The model}
The (0+1)-dimensional action proposed in Refs.~\cite{esterlis2019,hauck2020} is
	\be
	S = -\mu \sum_{i\sigma}\psi^\dag_{i\sigma}\psi_{i\sigma} + \frac 12 \int \dd\tau \sum_i \left[ (\partial_\tau \phi_i)^2 + \omega_0^2 \phi_i^2 \right]  + \frac 1N \int \dd\tau  \sum_{ijk,\sigma} g_{ij,k} \psi^\dag_{i\sigma}\psi_{j\sigma}\phi_k.
	\label{eq:S0+1}
	\ee
The first term is a chemical potential for the electrons; the second term describes a gapped boson (e.g., an optical phonon) with unit mass $m_\phi=1$ and bare frequency $\omega_0$; the last term is a ``Yukawa coupling" $g_{ij,k}$ between the electrons and bosons, which are taken to be complex-valued Gaussian-distributed random variables (with $g_{ij,k}^* = g_{ji,k}$, so that the corresponding Hamiltonian is Hermitian). Decomposing into real and imaginary parts as $g_{ij,k} = g'_{ij,k} + i g''_{ij,k}$, the distribution of the random couplings is determined by
	\begin{align}
	\overline{g'_{ij,k}g'_{i'j',k'}} &= \left(1-\frac\alpha 2 \right)  \overline g^2 \delta_{k,k'}(\delta_{ii'}\delta_{jj'} + \delta_{ij'}\delta_{ji'}), \\
	\overline{g''_{ij,k}g''_{i'j',k'}} &=\frac\alpha 2 \overline g^2 \delta_{k,k'}(\delta_{ii'}\delta_{jj'} - \delta_{ij'}\delta_{ji'}), \\
	\overline{g'_{ij,k} g''_{i'j',k'}} &= 0.
	\end{align}
The variance $\overline g^2$ measures the strength of the electron-boson interaction. The parameter $\alpha$, which lies in the range $0 \leq \alpha \leq 1$, is important for the superconducting properties: As we will elaborate upon below, a non-zero $\alpha$ leads to ``pair-breaking" effects, due to the fact that time-reversal symmetry is broken for individual realizations of the random couplings. In conventional superconductors, the effects of pair-breaking paramagnetic impurities were investigated in the pioneering work of Abrikosov and Gor'kov \cite{abrikosov1960}. The parameter $\alpha$ in the present model allows one to investigate the effects of similar pair-breaking disorder when the normal state is a NFL.

 The action \eqref{eq:S0+1} has three energy scales, $\mu$, $\omega_0$, and $\overline g^2/\omega_0^2$, and a natural dimensionless coupling parameter $g^2 = \overline g^2/\omega_0^3$. Taking $\omega_0$ as the unit of energy, the phase diagram of the system is thus determined by the parameters $\{g, \alpha,  \mu/\omega_0, M/N, T/\omega_0\}$, where $T$ is the temperature. We will mostly focus on the particle-hole symmetric point $\mu=0$ and the case of equal number of electron and boson flavors $M/N=1$. We comment briefly on the more general case at the end of this section and in Section~\ref{sec:holography} in relation to holographic superconductivity.

 A very similar model to \eqref{eq:S0+1} was proposed in Ref.~\cite{wang2020}, in which the bosonic degrees of freedom are matrix-valued $\phi_i \to \phi_{ij}$. While the normal state properties of that model are essentially the same as those of \eqref{eq:S0+1}, superconductivity appears at order $1/N$ in the case of Ref.~\cite{wang2020}.

\subsection{Solution in the large-$N$ limit}

\begin{figure}[t]%[h]
\includegraphics[width=0.7\textwidth]{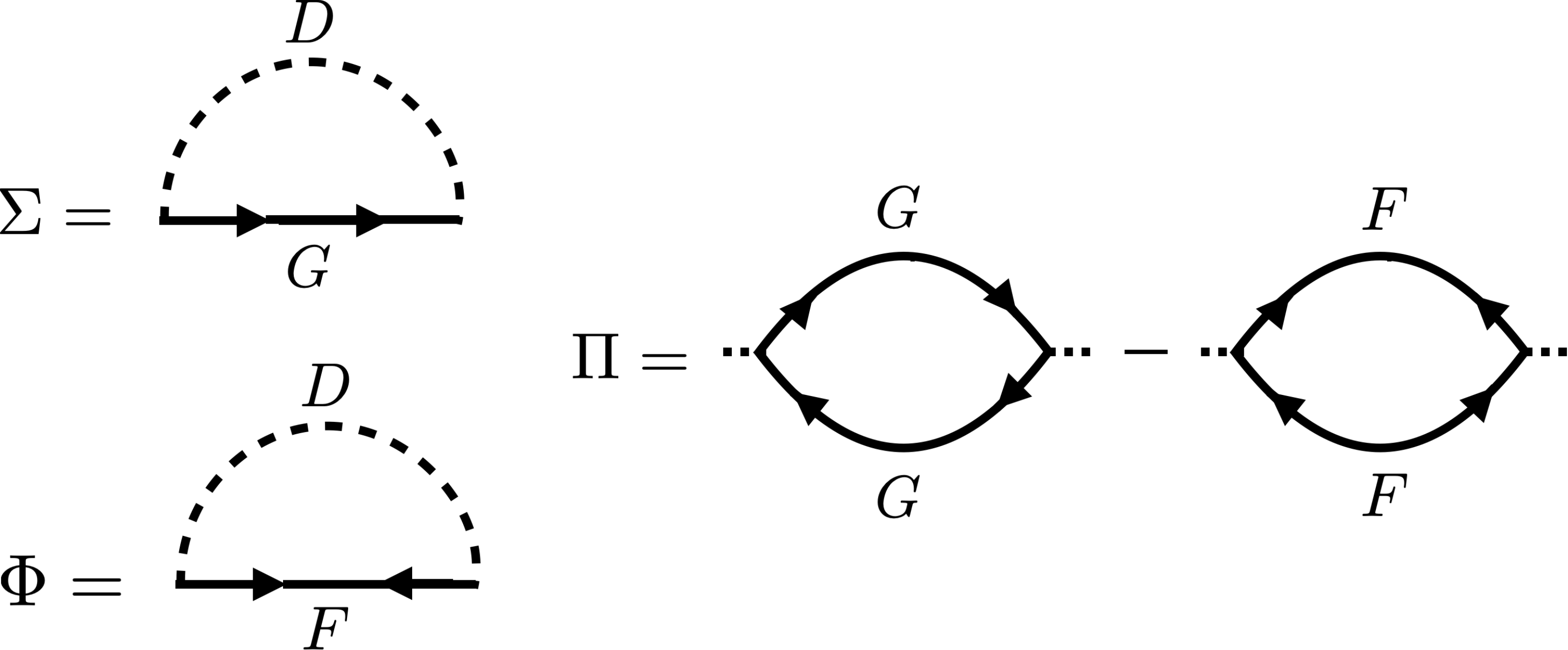}
\caption{Diagrammatic representation of the closed set of Eliashberg equations \eqref{eq:eliash_sig}-\eqref{eq:eliash_pi}. $\Sigma$ and $\Phi$ stand for the normal and anomalous electronic self-energies in Nambu space, respectively. $\Pi$ is the bosonic self energy. Solid (dashed)  lines denote dressed electron (boson) Green's functions.}
\label{fig:diags}
\end{figure}

The solution of the model \eqref{eq:S0+1} in the limit $N\to\infty$ proceeds in two steps:  Treating the problem in a path-integral formalism, the random couplings are eliminated in favor of the variance by the replica trick. The resulting effective action is then decoupled via a set of collective fields denoted $\{\Sigma, G, \Phi, F, \Pi,D\}$, which are bilocal in imaginary time; i.e., $\Sigma = \Sigma(\tau,\tau')$, etc. This generalizes the usual $G-\Sigma$ action that appears in SYK theories with purely fermionic degrees of freedom \cite{georges2001,maldacena2016}. Having expressed the effective action in terms of these bilocal fields, the $N \to\infty$ limit is taken, allowing the partition function to be evaluated in the saddle-point approximation. At the saddle-point level, the various fields become time-translation invariant, $\Sigma(\tau,\tau') \to \Sigma(\tau - \tau')$, etc. Details of the derivation can be found in \cite{esterlis2019,hauck2020}. Reliability of the replica trick and large-$N$ approximation will be discussed in Section \ref{sec:largeNcheck} below.

Remarkably, the saddle point equations reduce to a version of the familiar Eliashberg equations of superconductivity \cite{eliashberg1960} with a self-consistently calculated boson propagator,  the collective fields of the path integral playing the role of self-energies and Green's functions. The resulting equations, written on the Matsubara axis, are 
	\begin{align}
	\Sigma(i\omega_m) &= -\overline g^2 T \sum_{m'}D(i\omega_m - i\omega_{m'})G(i\omega_{m'}), \label{eq:eliash_sig}
	\\
	\Phi(i\omega_m) &= (1-\alpha)\overline g^2 T \sum_{m'}D(i\omega_m - i\omega_{m'})F(i\omega_{m'}), \label{eq:eliash_phi}
	\\
	\Pi(i\nu_m) &=  -2\overline g^2 T \sum_{m'} \left[G(i\omega_{m'} + i \nu_m) G(i\omega_{m'}) - (1-\alpha) F(i\omega_{m'} + i \nu_m) F(i\omega_{m'})\right].
    \label{eq:eliash_pi}
	\end{align}
Here bosonic and fermionic Matsubara frequencies are $\nu_m = 2\pi m T$ and $\omega_m = (2m+1)\pi T$, respectively. The inverse bosonic Green's function is $D^{-1}(i\nu_m) = \nu_m^2 + \omega_0^2 - \Pi(i\nu_m)$ and the inverse electronic Green's function in Nambu space is $\hat G^{-1}(i\omega_m) = i\omega_m \tau_0 - \hat\Sigma(i\omega_m)$, where $\tau_i$ are the Pauli matrices in Nambu space. We parametrize the fermion self-energy $\hat\Sigma$ in the usual way:
		\be
		\hat \Sigma(i\omega_m) = \Sigma(i\omega_m) \tau_0 + \Phi(i\omega_m) \tau_1,
		\ee
where $\Sigma$ and $\Phi$ are the ``normal" and ``anomalous" self-energies, respectively. The diagrammatic representation of these equations is shown in \textbf{Figure \ref{fig:diags}}. As we will describe later, the solution in higher spatial dimensions is formally similar. 

A key feature of the present model  for $M/N$ near unity is that, unlike other large-$N$ approaches to the problem of quantum critical metals,  here the electronic and bosonic degrees of freedom are both strongly dressed at the \textit{same order} in $1/N$. In this way one finds that the same boson-mediated interaction leads both to singular self-energy (i.e., non-Fermi liquid) effects \textit{and} superconductivity.

It is straightforward to extend the equations \eqref{eq:eliash_sig}-\eqref{eq:eliash_pi} to the real frequency axis \cite{schmalian1996}, where they can be solved to obtain the retarded Green's functions and self-energies. The retarded functions are important for extracting dynamical information about the system.

\subsubsection{Integral equations: Method of solution} 

Before presenting detailed results, we briefly comment on the methods of solution of the Eliashberg equations \eqref{eq:eliash_sig}-\eqref{eq:eliash_pi}. When the ground state is critical, the asymptotic low-frequency forms of the Green's functions and self-energies can be deduced by analytic methods, as described in detail in, e.g.,  Refs.~\cite{esterlis2019,wang2020,wang2020b,classen2021}. The procedure is standard in SYK-like models and involves making a power-law ansatz for the self-energies in Eqs.~\eqref{eq:eliash_sig}-\eqref{eq:eliash_pi}, then subsequently determining the exponent in the power-law self-consistently. The analytical approach is useful for obtaining singular power-law dependencies of various quantities on temperature and frequency, yielding directly the universal scaling laws governing the  properties of the system at criticality. 

To obtain more detailed information about the structure of the Green's functions, it is also useful to solve the equations directly by numerical methods. In addition to verifying analytically predicted scaling laws at criticality, the numerical approach provides insight into the system's behavior outside the quantum-critical regime. In practice, the equations are solved using an iterative scheme, with the computational procedure significantly accelerated by utilizing fast-Fourier transform (FFT) methods, as well as ``annealing" --  that is, using higher-temperature solutions as initial seeds for solving the equations at lower temperature. A ``hybrid frequency scale" method, which is especially useful for numerical analysis at low-$T$, has also been described in Ref.~\cite{abanov2020-II}.

\begin{figure}[t]
\includegraphics[width=\textwidth]{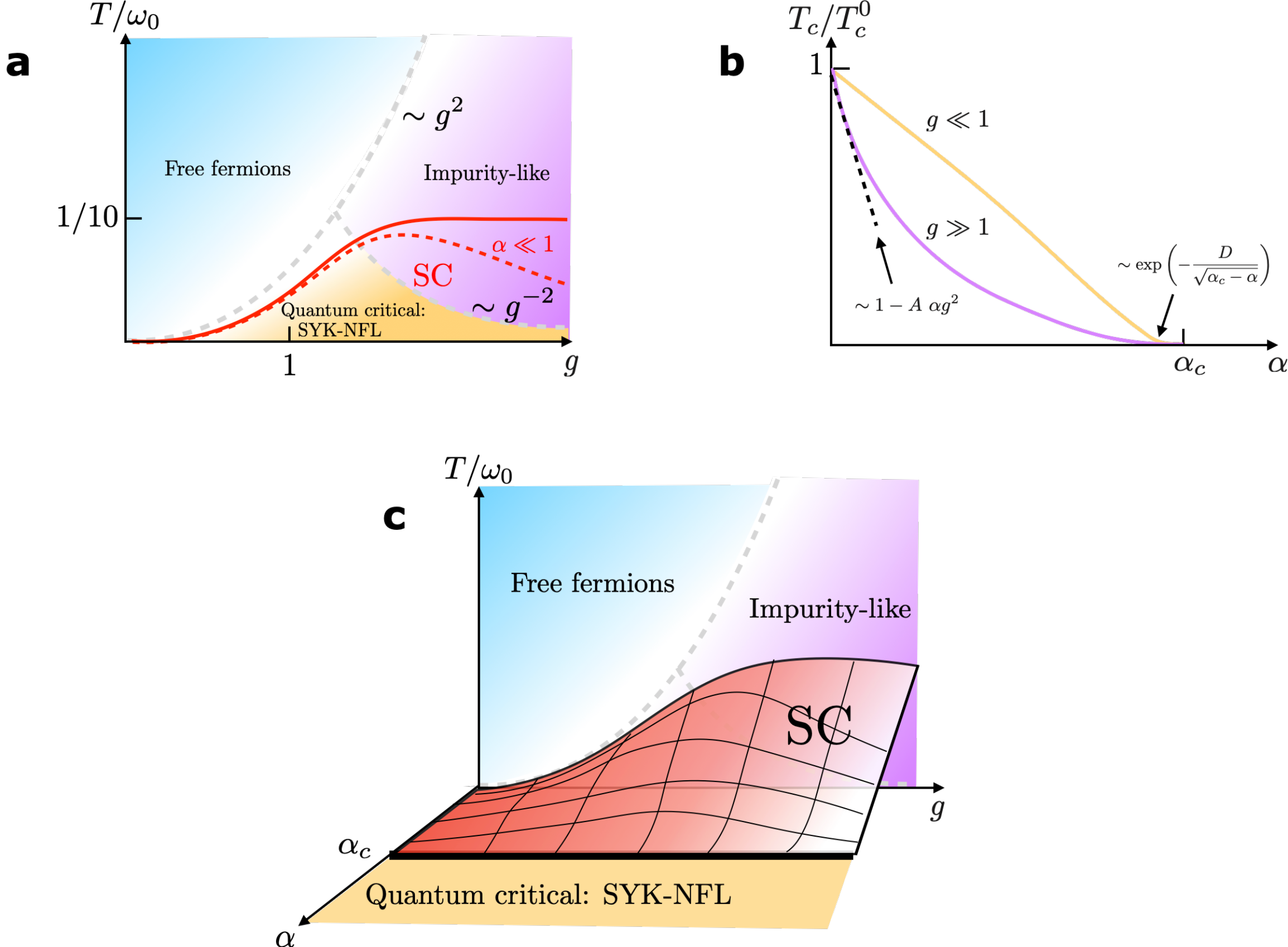}
\caption{Schematic phase diagram of the quantum dot YSYK model. Panel (a) is reproduced from Ref.~\cite{esterlis2019} and shows the crossovers between different regimes described in the text, as well as superconducting $T_c$ for values of the pair-breaking parameter $\alpha = 0$ and $\alpha \ll 1$. Panel (b) is reproduced from Ref.~\cite{hauck2020} and shows schematically the dependence of $T_c$ on $\alpha$ at weak and strong coupling, together with the BKT form for the vanishing of $T_c$ as $\alpha \to \alpha_c$ (see text). Panel (c) is the global phase diagram of the YSYK quantum dot as a function of $(g,\alpha,T)$.}
\label{fig:phase_diags}
\end{figure}

\subsection{Results} 

\subsubsection{Normal state properties}

We first describe the solution to Eqs.~\eqref{eq:eliash_sig}- \eqref{eq:eliash_pi} in the normal state, where the anomalous self-energy vanishes, $\Phi = 0$. (Equivalently, this describes the complete solution when $\alpha=1$, in which case superconductivity is entirely absent, as seen in Eq.~\eqref{eq:eliash_phi}.) Note that the pair-breaking parameter $\alpha$ drops out entirely from the normal state self-energy from Eq.~\ref{eq:eliash_sig}.

A unique feature of the YSYK model in the quantum dot limit is that, in the absence of superconductivity, the system self-tunes to criticality for \textit{any} value of the boson frequency $\omega_0$ and coupling $g$. That is, the renormalized boson frequency $\omega_r(T)$ vanishes as $T\to 0$ for any $\omega_0$ and $g$ according to 
	\be
	\omega_r^2(T) =  \omega_0^2 - \Pi(\nu = 0) \sim \left(\frac{T}{g^2}\right)^{4\Delta - 1},
    \label{eq:wr}
	\ee
where the exponent is found to be $\Delta \approx 0.420$ \footnote{The value of the $\Delta$ in general depends on the ratio of the number of boson to number of fermion flavors $M/N$ and varies in the range $\Delta \in (1/4,1/2)$.}. This ``self-tuning" to criticality is in stark contrast to the typical behavior of higher-dimensional systems, which require tuning a parameter to reach a critical point.

As $T \to 0$, the criticality of the boson leads to singular frequency dependence in both the electronic and bosonic self-energies, which take the low frequency form:
	\be
	\Pi(i\nu_m) \sim - \left| \frac{\nu_m}{g^2} \right|^{4\Delta -1 }, \quad \Sigma(i\omega_m) \sim -i \sgn(\omega_m) g^{4\Delta} |\omega_m|^{1-2\Delta},
    \label{eq:syknfl}
	\ee
where $\Delta$ is the same parameter as that entering Eq.~\eqref{eq:wr} and $\sim$ indicates suppression of numerical prefactors. As mentioned, we measure all energies and temperatures in units of the bare boson frequency $\omega_0$, while $g$ is dimensionless.
For frequencies $\omega \ll g^2$, the self-energy correction dominates the bare $i\omega$ term in the electron Green's function, implying ``non-Fermi liquid" behavior\footnote{In zero spatial dimensions, there is, of course, no Fermi surface. In this context, non-Fermi liquid refers to the dominance of self-energy corrections over the bare $i\omega$.}. The distinction between the NFL and the weakly interacting case is clearly demonstrated by considering the electronic spectral function $A(\omega) = -\im G_R(\omega)/\pi$ (where $G_R(\omega) = G(i\omega \to \omega + i\delta)$ is the retarded function): When $g=0$, we have simply $A(\omega) = \delta(\omega)$. On the other hand, in the NFL regime, the spectral function has a power-law form $A(\omega) \sim 1/|\omega|^{1-2\Delta}$, characteristic of a quantum critical system. The form of the electronic self-energy in Eq.~\eqref{eq:syknfl} is generic in SYK-like models, only differing in the numerical value of $\Delta$. We thus refer to the low-$T$ regime in which the electron and boson self-energies take the form \eqref{eq:syknfl}  as the ``SYK-NFL.''

In addition the quantum-critical SYK-NFL regime at lowest temperatures, a different sort of ``incoherent" electronic  regime emerges at strong-coupling $g > 1$ and intermediate temperatures. In this regime, which extends over the temperature window $g^{-2} \lesssim T \lesssim g^2$, the bosonic dynamics are undamped with Green's function $D^{-1}(i\nu_m) = \nu_m^2 + \omega_r^2$ and a renormalized frequency that is much smaller than the temperature, implying that the bosons are essentially classical:
	\be
	\omega_r^2(T) = \omega_0^2 - \Pi(\nu = 0) \sim \frac{T}{g^2}.
	\ee
The electrons thus effectively experience a static bosonic background, leading to a low-frequency electronic self-energy of the impurity type:
	\be
    \Sigma(i\omega_m) \approx  -i \sgn(\omega_m) \frac{\Omega_0}{2},
	\label{eq:imp}
	\ee
where the frequency scale is found to be $\Omega_0 = 16 g^2/3\pi$ \cite{esterlis2019}. In this regime, the impurity-like electron self-energy leads to a spectral function that is entirely smeared out, $A(\omega) \sim g^{-2}=\text{const.}$ at low frequencies. In this sense, the electrons are fully incoherent. We refer to this as the ``impurity-like" regime.
At large $g$, the impurity-like regime eventually crosses back over to the SYK-NFL regime described above for temperatures $T \lesssim g^{-2}$. The properties of the  normal state are summarized in the schematic phase diagram in \textbf{Figure \ref{fig:phase_diags}a}, displaying the qualitatively distinct SYK-NFL and impurity-like regimes.

\begin{figure}[t]
\includegraphics[width=\textwidth]{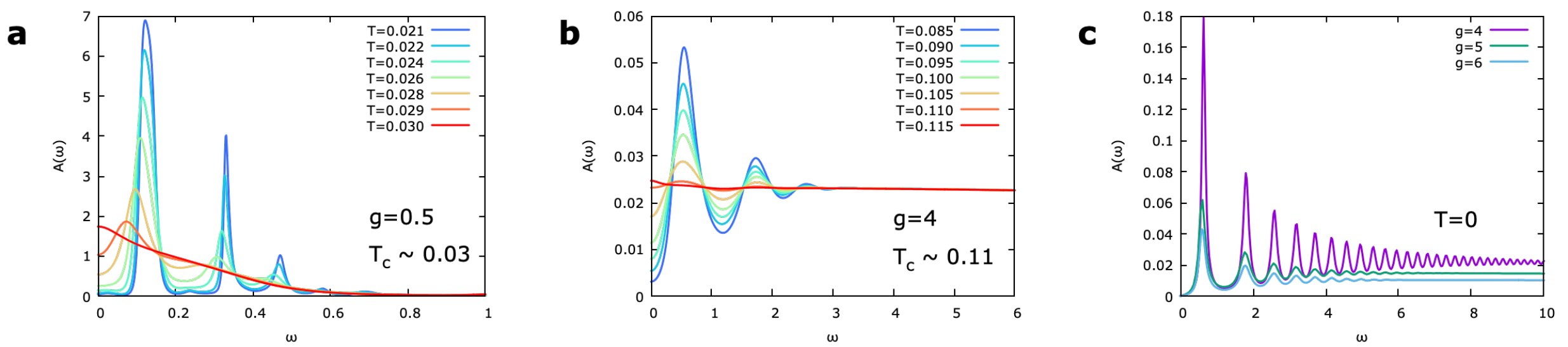}
\caption{Electronic spectral function $A(\omega)$ below $T_c$ for YSYK quantum dot, adapted from Ref.~\cite{esterlis2019}. Panels (a,b) show $A(\omega)$ for representative weak and strong values of the dimensionless coupling $g = 0.5$ and $g=4$, respectively. Panel (c) shows $A(\omega)$ at $T=0$ for representative strong coupling values $g=4,5,6$, demonstrating the emergence of the pronounced additional peaks in the strong coupling regime.}
\label{fig:sc_spectral_funcs}
\end{figure}

\subsubsection{Superconducting state} 
\label{sec:sc_dot}

Before going into details about specific superconducting properties of the YSYK quantum dot, it is instructive to make a connection to the universal properties of quantum-critical superconductors. This is accomplished by considering the linearized version of Eq.~\eqref{eq:eliash_phi} at the onset of superconductivity and inserting the scaling forms of the self-energies  \eqref{eq:syknfl}. We arrive at 
	\be
	\Phi(i\omega) =\frac{1-\alpha}{C_{\Delta}} \int \frac{\dd\omega'}{2\pi} \frac{\Phi(i\omega')}{|\omega - \omega'|^{4\Delta - 1} |\omega'|^{2-4\Delta}},
	\label{eq:universal_phi}
	\ee
where $C_{\Delta}=-8\cos\left(\pi\Delta\right)\sin^{3}\left(\pi\Delta\right)\Gamma\left(2\Delta\right)^{2}\Gamma\left(1-4\Delta\right)/\pi^{2}$ is a positive numerical coefficient that depends on the exponent $\Delta$. Here we have assumed the singular electron and boson self-energies dominate over the the bare propagators, and  that the onset temperature for superconductivity is sufficiently low that the Matsubara sums may be replaced by integrals over frequency. 
The power law structure of the integral equation immediately implies
a corresponding power-law solution \cite{chubukov2005}
\begin{equation}
    \Phi\left(\omega\right)\propto\left|\omega\right|^{-\frac{\gamma}{2}\pm i\beta},
    \label{eq:powerlawPhi}
\end{equation}
with, in general, complex exponent. The imaginary part of the exponent
vanishes at a critical value of the coupling constant as $\beta\propto\sqrt{\alpha_{c}-\alpha}$, where
 $\alpha_{c}$ is the value of the pair breaking parameter $\alpha$ where
$T_{c}$ vanishes~\cite{hauck2020} (see below). 

The gap equation in the form \eqref{eq:universal_phi} is ubiquitous across quantum-critical superconductors and has been investigated in great detail by Chubukov and collaborators in Refs.~\cite{wang2016,abanov2020-I,abanov2020-II,wu2020-III,wu2021-IV}, where it is referred to as a the $\gamma$-model  (with the relation $\gamma=4\Delta-1$). The distinction between different quantum critical systems appears in the precise value of the exponent $\Delta$. The gap equation is scale-invariant, with the temperature $T$ having entirely dropped out. A determination of $T_c$ therefore requires regularization, e.g., by retaining the bare $i\omega$ in the electronic Green's functions. In finite-dimensional systems the power-law behavior follows from the non-Fermi liquid normal state self energy $\Sigma\left(\epsilon\right)\sim\left|\epsilon\right|^{1-\gamma}$
and from the boson propagator $\int_{q_{\parallel}}D_{\boldsymbol{q}}\left(\epsilon\right)\sim\left|\epsilon\right|^{-\gamma},$
averaged over momenta tangential to the Fermi surface \cite{moon2010,abanov2020-I}.
Examples are $\gamma=1/2$ for $d=2$ spin-density wave instabilities \cite{abanov2001,abanov2001b}
and $\gamma\rightarrow0^{+}$, i.e. $\Sigma\left(\epsilon\right)\sim\epsilon\log$$\epsilon$,
for $d=3$ color superconductivity due to gluon exchange \cite{son1999}
or for three-dimensional spin fluctuations \cite{roussev2001,chubukov2005}. The limit 
$\gamma\rightarrow1^{-},$ i.e. $\Sigma\left(\epsilon\right)\sim\log$$\epsilon$,
follows for $d=3$ massless bosons at very strong coupling \cite{chubukov2005}.
$\gamma=1/3$ describes an Ising ferromagnetic QCP,  composite fermions at half-filled Landau
levels \cite{bonesteel1996}, emergent gauge fields \cite{lee1989gauge,polchinski1994low}
or nematic transitions \cite{klein2020} in two dimensions. Hence, while the quantum dot description discussed here is at first glance a rather artificial and simplified model, it shares crucial aspects with theories for quantum critical superconductivity that have been analyzed in a range of two- and three-dimensional theories near a quantum-critical point. 

We now turn to a more quantitative analysis of superconductivity of the YSYK quantum dot, which requires solving all three equations \eqref{eq:eliash_sig}, \eqref{eq:eliash_phi}, and \eqref{eq:eliash_pi} simultaneously. When $\alpha = 0$, the first  thing to observe is that the ground state of the model is a spin singlet superconductor for all $g$. At weak coupling $g \lesssim 1$, the transition temperature has the dependence $T_c/\omega_0 \sim g^2$ \cite{esterlis2019}. The scale for superconductivity coincides with the crossover into the SYK-NFL, and superconductivity thus preempts a broad NFL regime. We emphasize that the transition temperature has a power-law dependence on the coupling and is therefore parametrically larger than the exponentially small $T_c$ of a BCS superconductor. This can be explicitly shown if one adds a small single-particle inter-dot hopping that yields a low-$T$ Fermi liquid with BCS behavior~\cite{valentinis2023_lett,valentinis2023}.  At strong coupling the superconducting transition temperature saturates to a $g$-independent value $T_c \approx 0.1\omega_0$ (see \textbf{Figure \ref{fig:phase_diags}a}). The saturation behavior is a version of Anderson's theorem \cite{andeson1959,abrikosov1959,abrikosov1959b,abrikosov1960,potter2011,kang2016,millis1988,abanov2008}: at large $g$, superconductivity emerges from the impurity-like regime, where the boson frequency is softened and the boson behaves essentially like static, non-magnetic disorder. Anderson's theorem implies that, despite the dramatic softening, this ``effective disorder" cannot suppress the superconducting $T_c$. This may be seen more directly by considering the linearized gap equation for the frequency-dependent gap function $\Delta(i\omega_m) = i\omega_m\Phi(i\omega_m) / [i\omega_m-\Sigma(i\omega_m)]$:
    \be
    \Delta(i\omega_m) = g^2 T_c \sum_{m'} \frac{D(i\omega_m -i\omega_{m'})}{\omega_{m'}+i\Sigma(i\omega_{m'})} \left[ (1-\alpha)\frac{\Delta(i\omega_{m'})}{\omega_{m'}}  - \frac{\Delta(i\omega_m)}{\omega_m}\right],
    \label{eq:gapeq}
    \ee
from which one sees immediately that, when $\alpha=0$, the zero Matsubara frequency transfer term (i.e., the term in the sum $m' = m$), associated with static fluctuations, drops out. 

Focusing still on $\alpha = 0$, the nature of the superconducting state itself is different in the weak and strong coupling regimes, as may be observed from the electronic spectral function $A(\omega)$ below $T_c$. At weak coupling $g\lesssim 1$, where superconductivity preempts a broad NFL regime, $A(\omega)$ displays a sharp Bogoliubov quasi-particle peak, the position of which moves toward $\omega=0$ as $T \to T_c$ from below; i.e., the spectral function exhibits a ``gap closing" behavior; see \textbf{Figure \ref{fig:sc_spectral_funcs}a}. The situation is qualitatively different at strong coupling $g \gtrsim 1$, when superconductivity emerges deep in the impurity-like regime where the electrons are highly incoherent. In this case $A(\omega)$ exhibits a ``gap-filling" behavior as $T_c$ is approached from below, with the position of the quasi-particle peak being essentially independent of $T$ (\textbf{Figure \ref{fig:sc_spectral_funcs}b}). The spectral function also exhibits a significant amount of structure at high frequencies in the form of sharp peaks, which are most pronounced at $T=0$ (\textbf{Figure \ref{fig:sc_spectral_funcs}c}). These features in the strong-coupling limit of the Eliashberg equations were first observed by Combescot in \cite{combescot1995}, where they were interpreted as ``self-trapping" states of an excited electron in the self-consistent pairing potential of the other electrons, similar to satellite peaks associated with polaron formation. This suggests a picture of the superconducting state in which the Cooper pairs are not simply a dilute gas, as in a conventional BCS superconductor, but rather form a strongly interacting pair fluid. Yet another remnant of the normal state electronic incoherence in the superconducting state at strong coupling is the reduced weight of the Bogoliugbov quasi-particle peak, which follows $Z_B \sim g^{-2}$ \cite{esterlis2019}. This property is directly related to the incoherent electronic normal state where $A(\omega) \sim g^{-2}$, since the weight of the peak can be estimated as being transferred from energies below the gap: $Z_B \approx \int_0^\Delta \dd\omega A(\omega) \sim g^{-2}$.

It is important to note that the quantum dot theory of superconductivity is, of course, a mean field theory. Not unlike models of Heisenberg spin systems with all-to-all couplings that display a phase transition but no spin-wave stiffness, the definition of a superfluid  stiffness in the quantum dot model is not possible. This can, however,  be rectified in higher dimensional extensions~\cite{valentinis2023_lett,valentinis2023,li2024}, where  a relationship between stiffness and small Bogoliubov weight was identified at large coupling.

The ground state of the YSYK quantum dot remains superconducting 
when the pair-breaking parameter $\alpha$ lies in the range $0 \leq \alpha \leq \alpha_c$, where $\alpha_c \approx 0.62$ is a critical $g$-independent value beyond which superconductivity disappears entirely \cite{hauck2020}. The critical $\alpha_c$ can be determined by analyzing the gap equation \eqref{eq:universal_phi}, from which one also finds that $T_c$ vanishes with the a Berezinskii–Kosterlitz–Thouless (BKT) form $T_c \sim T^* \exp\left(-D/\sqrt{\alpha_c - \alpha}\right)$ \cite{hauck2020}, where $T^* \sim \omega_0 \min(g^2,g^{-2})$ is the energy scale below which the SYK-like scaling \eqref{eq:syknfl} sets in. This BKT-like behavior has been argued to be a generic feature of transitions from a conformal to non-conformal phase \cite{kaplan2009} and, in the case of the YSYK quantum dot, is a reflection of the emergent conformal symmetry of the SYK-NFL \eqref{eq:syknfl}. It is closely related to the vanishing of the imaginary part  $\beta$ of the exponent in Eq.~\eqref{eq:powerlawPhi}.

For $\alpha < \alpha_c$ and at weak coupling $g\ll 1$, the pair breaking effects are relatively weak (see \textbf{Figure \ref{fig:phase_diags}b}). In contrast, the superconductivity at strong coupling $g \gg 1$ is much more fragile, with $T_c \approx T_c^0(1-A \alpha g^2)$ \cite{hauck2020}, where $T_c^0$ is the transition temperature for $\alpha = 0$ (no pair breaking) and the numerical coefficient $A \approx 1.1$ (\textbf{Figure \ref{fig:phase_diags}b}). At large $g$, superconductivity is thus significantly suppressed for $\alpha > \alpha^*$, where  $\alpha^*\sim g^{-2} \ll \alpha_c$. The scale for $\alpha^*$ is naturally identified with the Bogoliubov quasiparticle weight $Z_B \sim g^{-2}$, which leads to the important conclusion that a reduced quasiparticle weight results in a fragile superconducting state that is highly sensitive to pair-breaking disorder. 
A schematic phase diagram as a function $(g,\alpha,T)$ summarizing the evolution between the SYK-NFL and impurity-like normal states, together with the superconducting $T_c$, is presented in \textbf{Figure \ref{fig:phase_diags}c}.

\subsubsection{Deformations of the model} Here we briefly summarize important changes in the properties of the model \ref{eq:S0+1} when $\mu\neq 0$ \cite{wang2020b}, and when the ratio of the number of boson to fermion flavors $M/N$ is varied \cite{classen2021}. 

Focusing on the normal-state properties at weak-coupling, it was found in Ref.~\cite{wang2020b} that there is a first-order transition with increasing $\mu$ (which also tunes the system away from particle-hole symmetry) from an SYK-like NFL regime like that described above to an insulating state at large $\mu$. In the small $\mu$ SYK-like NFL regime, the self-energies retain their power-law form as in \eqref{eq:syknfl}, with the electronic self-energy becoming asymmetric about $\omega = 0 $ and an associated asymmetry in the electronic spectral function. Interestingly, the exponent $\Delta$ is continuously varying with $\mu$ in this regime. When the the number of bosons greatly exceeds the number of fermions, $M/N \to \infty$, the first-order transition to the insulator was shown to become continuous \cite{wang2020b}. A subsequent DQMC study \cite{wang2021_dqmc} investigated the interplay between the NFL, insulating, and superconducting states, finding that the NFL-insulator transition is ultimately masked by a superconducting phase (although the details of this also depend on the ratio $M/N$ \cite{wang2021_dqmc}).

It was shown in Ref.~\cite{classen2021} that a rich phase diagram also emerges with varying $M/N$. The crossover between SYK-NFL and impurity-like regimes described above can be tuned with varying $M/N$ and, in the limit $M \gg N$, entirely new regimes appear, characterized by a more complex interplay between the quantum and thermal fluctuations in the normal state \cite{classen2021}. At weak coupling the effect of a differing number of electron and boson flavors on the superconducting $T_c$ is small, while at strong coupling the saturation value of the transition temperature is $T_c \sim \sqrt{M/N} \times \omega_0$.

\subsection{Reliability of replica trick and large-$N$}
\label{sec:largeNcheck}
The approximations that go into deriving the Eliashberg equations \ref{eq:eliash_sig}-\ref{eq:eliash_pi} from the original model \eqref{eq:S0+1} are the replica-trick treatment of the random couplings, with the subsequent restriction to replica-diagonal solutions, and the large-$N$ limit. In much of the parameter space, the validity of these approximations has been supported by numerically exact DQMC simulations \cite{pan2021,wang2021_dqmc}, which studied a particular deformation of the model originally proposed in Ref.~\cite{wang2020}. These simulations were carried out for finite $N$, for many realization of the random couplings, and then averaged. With increasing $N$, the DQMC results were found to approach the large-$N$ predictions  smoothly. In certain variants of the YSYK model, however, the large-$N$ approach is found to break down when compared with the exact DMQC simulations \cite{pan2021,esterlis_unpublished}. This is found to occur, for example, in the model \eqref{eq:S0+1} at sufficiently strong coupling. 
 In this regime, the DQMC simulations show signatures of ``glassy" behavior, which may be possible to explain analytically via a replica-symmetry breaking transition \cite{pan2021}. The emergence of a glass phase--or ``glassy superconductivity"--is an intriguing possibility that warrants further investigation. It is also important to note that taking the large-$N$ limit of the YSYK quantum dot suppresses certain physical effects present at finite $N$, such as polaron formation \cite{mahan2013}. 

 \subsection{Significance and Takeaways} 
In this section, we have described an exactly solvable large-$N$ model of electrons and bosons interacting via random, all-to-all couplings. Despite its simplicity, this model exhibits a number of non-trivial behaviors: The exact solution reveals two distinct strongly-interacting regimes: a ``SYK-like" quantum critical regime with singular power-law self-energies \eqref{eq:syknfl}, and an ``impurity-like" regime at strong-coupling \eqref{eq:imp}, in which the electronic spectral function is broad. When the pair-breaking effects, captured by the parameter $\alpha$, are sufficiently weak, the ground state of the system is superconducting, independent of the coupling strength. The nature of the superconductor is markedly different, however, at weak and strong coupling. At weak coupling, superconductivity preempts a broad regime of quantum critical scaling in the normal state, explicitly demonstrating the tension between the tendency of quantum critical fluctuations to destabilize the FL while also promoting electron pairing. At strong coupling, superconductivity emerges from fully incoherent electrons, which have been smeared out due to their strong coupling to nearly-classical bosonic fluctuations. The resulting superconducting state appears to be best described as a strongly interacting Cooper pair fluid. This state is fragile, with a pronounced sensitivity to pair-breaking disorder. Given the close relation between the linearized gap equation of the quantum dot model and that found in compressible, finite-dimensional systems, as well as the documented reliability of the Eliashberg approach to problems of strong-coupling superconductivity \cite{wang2017,klein2020,xu2020,chubukov2020,chowdhury2020_eliash}, it is reasonable to expect that the conclusions drawn here regarding the interplay between normal-state incoherence and superconductivity extend beyond the simplified YSYK quantum dot model from which they were obtained.

\section{Higher dimensions} 
\label{sec:higher_dim}

We now turn to higher-dimensional generalizations of the model \eqref{eq:S0+1}, which have recently been proposed and analyzed as part of efforts to construct a more realistic theory of quantum critical metals and superconductors. Here, we focus on the approach adopted in Refs.~\cite{esterlis2021,guo2022,patel2023,guo2024_cyclotron,li2024,aldape2022}, omitting various technical details that are similar to those in the quantum dot case.  A different lattice generalization was explored in \cite{valentinis2023_lett,valentinis2023},  where the electronic hopping $t_{ij}(x,x')$ between nearest-neighbor sites $x$ and $x'$ was taken to be random in flavor space (indices $i$ and $j$);  we will comment on some of those results below. In this section, our primary focus will be on the \textit{normal-state} properties of the system. The superconducting properties in higher dimensions are broadly similar to those of the quantum dot problem--reducing to a version of Eq.~\eqref{eq:universal_phi}--with the main distinction arising in the normal state due to the presence of a Fermi surface. 

\subsection{The model}
In $d$ spatial dimensions and in the long-wavelength limit, the action we will consider is 
	\be
	\begin{aligned}
	S &= \int \dd\tau \int \dd^d x \sum_j \psi^\dag_{j\sigma}\left[\partial_\tau + \varepsilon(i\nabla) - \mu \right]\psi_{j\sigma} \\
	&\qquad + \frac 12 \int \dd\tau \int \dd^d x \sum_i \left[ (\partial_\tau\phi_i)^2 + c^2(\nabla \phi_i)^2 + \omega_0^2 \phi_i^2 \right] \\
	&\qquad + \frac 1N \int \dd\tau  \int \dd^d x \sum_{ijk,\sigma} \left[ g_{ij,k}  + g'_{ijk}(x)\right] \psi^\dag_{i\sigma}\mathcal \psi_{j\sigma}\phi_k.
	\end{aligned}
	\label{eq:Sd+1}
	\ee
This is a natural generalization of the action \eqref{eq:S0+1}, in which the electrons have a dispersion $\varepsilon(k)$ and bosons have a long-wavelength dispersion law $\Omega^2(q) = c^2 q^2 + \omega_0^2$. The couplings $g_{ij,k}$ and $g'_{ij,k}(x)$ are again taken to be random variables in the flavor space with zero mean, and variances $g^2$ and $g'^2$. 
The crucial difference is that $g_{ij,k}$ is infinitely correlated in space and preserves translational invariance, whereas $g'_{ijk}(x)$ is uncorrelated between different sites. Consequently, the randomness in $g_{ij,k}$ serves primarily as a theoretical tool for controlled calculations in non-disordered systems, while $g'_{ijk}(x)$ captures genuine impurity effects through a spatially varying electron-boson interaction.

The model \eqref{eq:Sd+1} can also be viewed as a particular large-$N$ deformation of the standard Hertz-Millis \cite{hertz1976,millis1988} type theories of metallic quantum criticality. The fermions $\psi$ are the itinerant metallic degrees of freedom, while $\phi$ represents fluctuations of an appropriate order parameter. In general, the nature of the ordering transition is reflected in the type of bosonic field used (scalar, vector, etc.) and the spatial and spin structure of the fermion-boson coupling. In this context, the action in \eqref{eq:Sd+1} gives the low-energy description of a uniform ($\mathbf{Q}=0$) ordering transition; e.g., an Ising-nematic transition. (For simplicity, however,  the appropriate form factors have not been included in the Yukawa interaction written above, as they are not expected, in the absence of coupling to phonon modes \cite{paul2017lattice}, to change any qualitative physics associated with the non-Fermi liquid effects or superconductivity.) The case of an ordering transition with $\mathbf{Q} \neq 0$; e.g., a charge or spin-density wave transition, is amenable to a similar YSYK large-$N$ analysis.

We also comment on the motivations for including both the homogeneous $g$ coupling and the spatially disordered $g'$ coupling. When $g'=0$, the model \eqref{eq:Sd+1} describes the properties of a clean metallic quantum critical point and, as we will describe below, leads to a NFL normal state in the large-$N$ limit. However, the clean system fails to reproduce the transport properties that are the hallmark of strange metals phases -- in particular the $T$-linear resistivity -- as it possesses an infinite dc conductivity \cite{guo2022}.  While conventional elastic impurity scattering is readily included, experimental results such as those in \cite{michon2023} point to the importance of an \textit{inelastic} scattering mechanism in the strange metal regime, as evidenced from the strongly frequency-dependent scattering rate. As we will elaborate upon below, the $g'$ coupling leads to inelastic and momentum non-conserving scattering, while also maintaining the solvability of the model in the large-$N$ limit. 
In connection with more familiar types of quenched disorder, the disordered Yukawa coupling can be understood as a simplified approximation to ``random mass" disorder for the boson, representing spatial inhomogeneity in the location of the quantum critical point \cite{patel2024_localized-bosons}. So long as the bosonic eigenmodes are not localized by the disorder -- which occurs in certain regions of the phase diagram at sufficiently low $T$ -- the random-mass and random-coupling models behave similarly \cite{patel2024_dqmc,patel2024_localized-bosons}. 

The model \eqref{eq:Sd+1} has three important energy scales: the electronic bandwidth $W$ (we assume the Fermi energy $\varepsilon_F$ is of order the bandwidth), a bosonic cutoff energy scale $W_b$, and an interaction energy $g^2/W_b^2$ ($g$-model) or $g'^2/W_b^2$ ($g'$-model).  We will be primarily interested in the regime where the interaction energy is smaller than the bandwidth; at stronger coupling the behavior crosses over to that of the quantum dot model described above. Another important dimensionless parameter is the ratio $c/v_F$ where $v_F$ is the Fermi velocity; we will focus on the regime $c \sim v_F$.  Unlike the quantum dot, in higher dimensions the system must be tuned to the quantum critical point by varying $\omega_0^2$, the bare boson mass\footnote{In practice, the QCP can also be accessed by imposing a ``fixed length'' constraint on the boson,  introduced by adding a quartic self-interaction term to the action $S_u = (u/2N) \int_{\tau, x} ~ \left(\sum_i \phi_i \phi_i - N/\gamma\right)^2$, and taking the limit $u\to \infty$. In the large-$N$ limit, this yields the constraint $D(r = 0, \tau =0)=1/\gamma$, with $\gamma$ used to access the QCP \cite{esterlis2021,li2024}.}.

\begin{figure}[t]
\includegraphics[width=\textwidth]{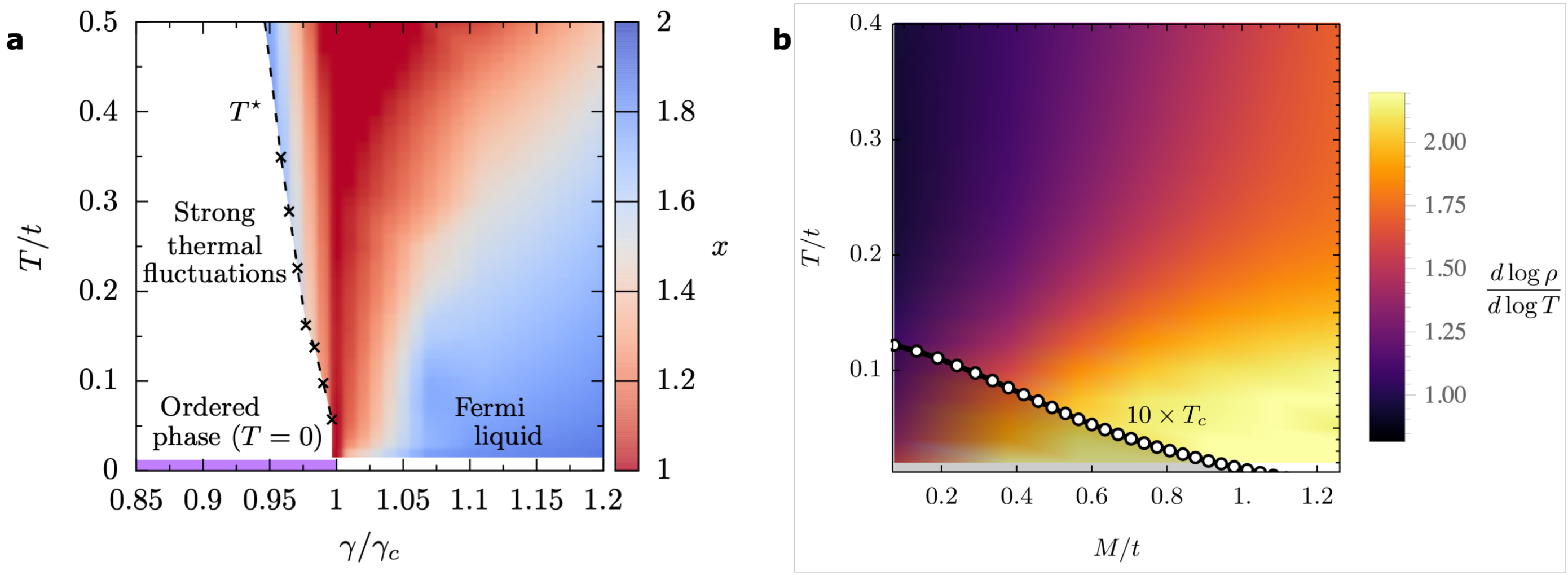}
\caption{(a) Phase diagram of the 2D YSYK model with coupling $g'=0$ and $g=2 t^{3/2}$ as a function of the tuning parameter $\gamma$ used to access the QCP at $\gamma_c$ . Color scale indicates the exponent $x$ with which the renormalized boson mass approaches its $T=0$ value: $M^2(T) - M^2(T=0) \sim T^x$, demonstrating the emergence of a quantum critical fan. The dashed line $T^*$ denotes the onset of strong thermal fluctuations of the boson, below which the renormalized mass becomes exponentially small and the system is nearly ordered \cite{esterlis2021}. (b) Phase diagram of the $g'$-model with $g' = 2t^{3/2}$ and $g=0$ as a function of the renormalized boson mass $M(T=0)$ and $T$; $M(T=0) = 0$ at the QCP.  Color scale indicates the approximate exponent in the $T$-dependence of resistivity; see Fig.~\ref{fig:gp_rho+sig}a for line cuts of $\rho(T)$. Superconducting $T_c$ is also shown.
Panel (a) is adapted from Ref.~\cite{esterlis2021} and panel (b) is adapted from Ref.~\cite{li2024}.}
\label{fig:g+gp_phase_diags}
\end{figure}

\subsection{Results}

In this section we will summarize key results in the analysis of the action \eqref{eq:Sd+1}, focusing, for simplicity, separately on the translationally invariant case ($g'=0$) and purely disordered case ($g=0$). Below we refer to these as the ``$g$-model" and ``$g'$-model", respectively. Properties of the model with $g$ and $g'$ both non-zero have been discussed in \cite{esterlis2021,guo2022,patel2023}. We specialize to spatial dimension $d=2$, which is the most interesting case in connection to real materials.

\subsubsection{$g$-model} The integral equations that solve the $g$-model in the large-$N$ limit are formally identical to those in (0+1)-dimensional \eqref{eq:eliash_sig}-\eqref{eq:eliash_pi} but with wave-vector dependent Green's functions and self-energies:
	\begin{align}
	\Sigma(\bfk,i\omega_m) &= -g^2 T \sum_{m'}\int_{\bfk'} D(\bfk-\bfk',i\omega_m - i\omega_{m'})G(\bfk',i\omega_{m'}), \label{eq:eliash_sig2}
	\\
	\Pi(\bfq,i\nu_m) &=  -2g^2 T \sum_{m'}\int_k G(\bfk+\bfq,i\omega_{m'} + i \nu_m) G(\bfk,i\omega_{m'}).
    \label{eq:eliash_pi2}
	\end{align}
Though we do not pause to include it here, when the couplings $g_{ij,k}$ are real there is an additional equation for the anomalous self-energy as in the quantum dot case \eqref{eq:eliash_phi}. 

At zero temperature and precisely at the quantum critical point, analytic expressions can be obtained for the boson and fermion self-energies in the case of an isotropic quadratic band with constant density of states $\mathcal N$ and Fermi energy $\varepsilon_F$:
	\be
	\Pi(\bfq, i\nu_m) = -2g^2 \mathcal N \frac{|\nu_m|}{v_F q}, \quad \Sigma(\bfk_F, i\omega_m) \sim -i\sgn(\omega_m)\left(\frac{g^4}{\varepsilon_F c^4}\right)^{1/3}|\omega_m|^{2/3}.
	\label{eq:z3nfl}
	\ee
The electronic self-energy is evaluated on the Fermi surface at $\bfk_F$ and $\sim$ indicates suppression of numerical prefactors. The boson self-energy has the usual Landau-damping form, leading to $z=3$ dynamical exponent; i.e. $\omega \sim i q^3$ overdamped dynamics when analytically continued to the real axis.  At low frequencies, the singular electron self-energy dominates the bare $i\omega$ in the electron Green's function, yielding a NFL state. 

The above results are well-known for the Ising-nematic QCP, having been derived earlier by a variety of methods. The virtue of the YSYK approach is that it offers a large-$N$ limit in which the results are exact, and the boson and electron renormalizations occur at the same order in $1/N$. Moreover, because the integral equations \eqref{eq:eliash_sig2} and \eqref{eq:eliash_pi2} are also valid at non-zero $T$ above the critical point, they may be used to diagnose the effects of thermal fluctuations on the quantum critical properties, as well as away from the critical region to obtain a more complete picture of the phase diagram. This includes the crossovers between different regimes, which are challenging to address analytically. 

An analysis of the complete phase diagram was carried out in Ref.~\cite{esterlis2021}, where the model \eqref{eq:Sd+1} was discretized on a 2d square lattice and Eqs.~\eqref{eq:eliash_sig2} and \eqref{eq:eliash_pi2} were solved numerically. The resulting phase diagram is reproduced in \textbf{Figure \ref{fig:g+gp_phase_diags}a}. A quantum critical fan emerges above the QCP, as indicated by the exponent $x$ which governs the low-$T$ behavior of the boson mass: $M^2(T) - M^2(T=0) \sim T^x$. In the quantum critical fan, the expected behavior is $M^2 \sim T \log(1/T)$, i.e., an exponent $x$ close to unity. The fan is flanked by a low-$T$ Fermi liquid regime and a (nearly) ordered phase\footnote{The $O(N)$ symmetry of the disorder (in flavor space) averaged theory forbids a finite-$T$ transition, which is instead replaced by a regime of strong thermal fluctuations, in which the boson acquires an exponentially small mass below a temperature scale $T^*$. This, in turn, leads to impurity-like behavior of the electronic self-energy, reminiscent of the impurity regime in the (0+1)-dimensional model.}. 

\begin{figure}[t]
\includegraphics[width=\textwidth]{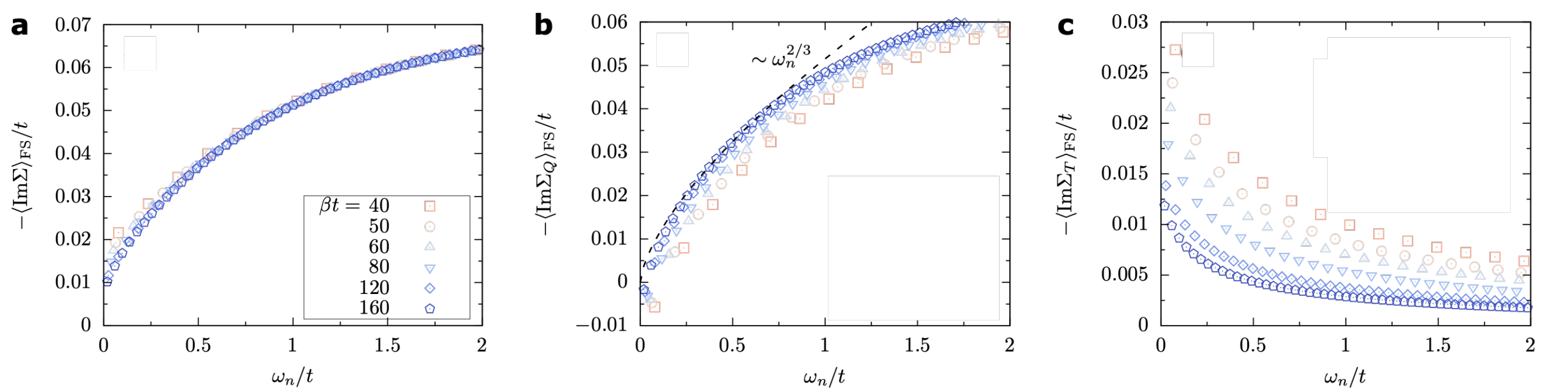}
\caption{Imaginary part of the electronic self-energy averaged over the Fermi-surface (FS) for the $g$-model with $g=2t^{3/2}$ at the QCP, for inverse temperatures $\beta = 1/T$ indicated in the inset of (a). Panel (a) shows the full self-energy, while panels (b) and (c) show, respectively, the quantum and thermal contributions to the self-energy as defined by Eq.~\eqref{eq:eliash_sigQ} and \eqref{eq:eliash_sigT}. Panel (b) demonstrates the expected scaling $-\im\Sigma_Q \sim |\omega|^{2/3}$ \eqref{eq:eliash_sig2}. Figures reproduced from Ref.~\cite{esterlis2021}.}
\label{fig:sig_decomp}
\end{figure}

Within the quantum critical fan, the full numerical solution shows that thermal fluctuations of the boson -- corresponding to the zero bosonic Matsubara frequency term in Eq.~\eqref{eq:eliash_sig2} -- mask the quantum critical scaling of Eq.~\eqref{eq:z3nfl} (\textbf{Figure \ref{fig:sig_decomp}a}). Nevertheless, the expected scaling behavior may be exposed by decomposing the  electronic self-energy into ``quantum" ($Q$) and ``thermal" ($T$) contributions $\Sigma(\bfk,i\omega_m) =\Sigma_Q(\bfk,i\omega_m) + \Sigma_T(\bfk,i\omega_m)$ \cite{dellanna2006,damia2020,damia2021,klein2020,xu2020}, defined according to 
	\begin{align}
	\Sigma_Q(\bfk,i\omega_m) &= -g^2 T \sum_{m' \neq m}\int_{\bfk'} D(\bfk-\bfk',i\omega_m - i\omega_{m'})G(\bfk',i\omega_{m'}), \label{eq:eliash_sigQ}
	\\
    \Sigma_T(\bfk,i\omega_m) &= -g^2 T \int_{\bfk'} D(\bfk-\bfk',i\nu_m = 0)G(\bfk',i\omega_m). \label{eq:eliash_sigT}
	\end{align}
The separate contributions are shown in \textbf{Figure \ref{fig:sig_decomp}b,c}. Scaling is spoiled by the slow decay of the thermal part at low temperatures and frequencies $\im \Sigma_T \sim -i \sgn(\omega_m)(g/c) \times \sqrt{  T \log(1/T)}$ \cite{klein2020,esterlis2021} (see \textbf{Figure \ref{fig:sig_decomp}c}). Only after isolating $\Sigma_Q$ is the expected quantum-critical scaling observed  (\textbf{Figure \ref{fig:sig_decomp}b}). We have included this analysis in order to highlight some of the subtleties involved in extracting scaling behavior of electronic observables near a (clean) metallic quantum critical point \cite{klein2020,xu2020}. 

In the quantum critical regime, the pairing problem reduces to the universal quantum-critical form \eqref{eq:universal_phi}, with the particular value $\Delta = 1/3$ and a modified coefficient \cite{esterlis2021}. We emphasize that this equation now emerges as an exact result in the large-$N$ limit. Estimating the transition temperature yields $T_{c} \sim g^4$, which has the same parametric dependence as the NFL scale below which the electronic self-energy \eqref{eq:eliash_sig2} dominates over the bare $i\omega$ in the Green's function. Thus one does not expect, in general, a broad NFL regime above $T_c$. 

\begin{figure}[t]
\includegraphics[width=\textwidth]{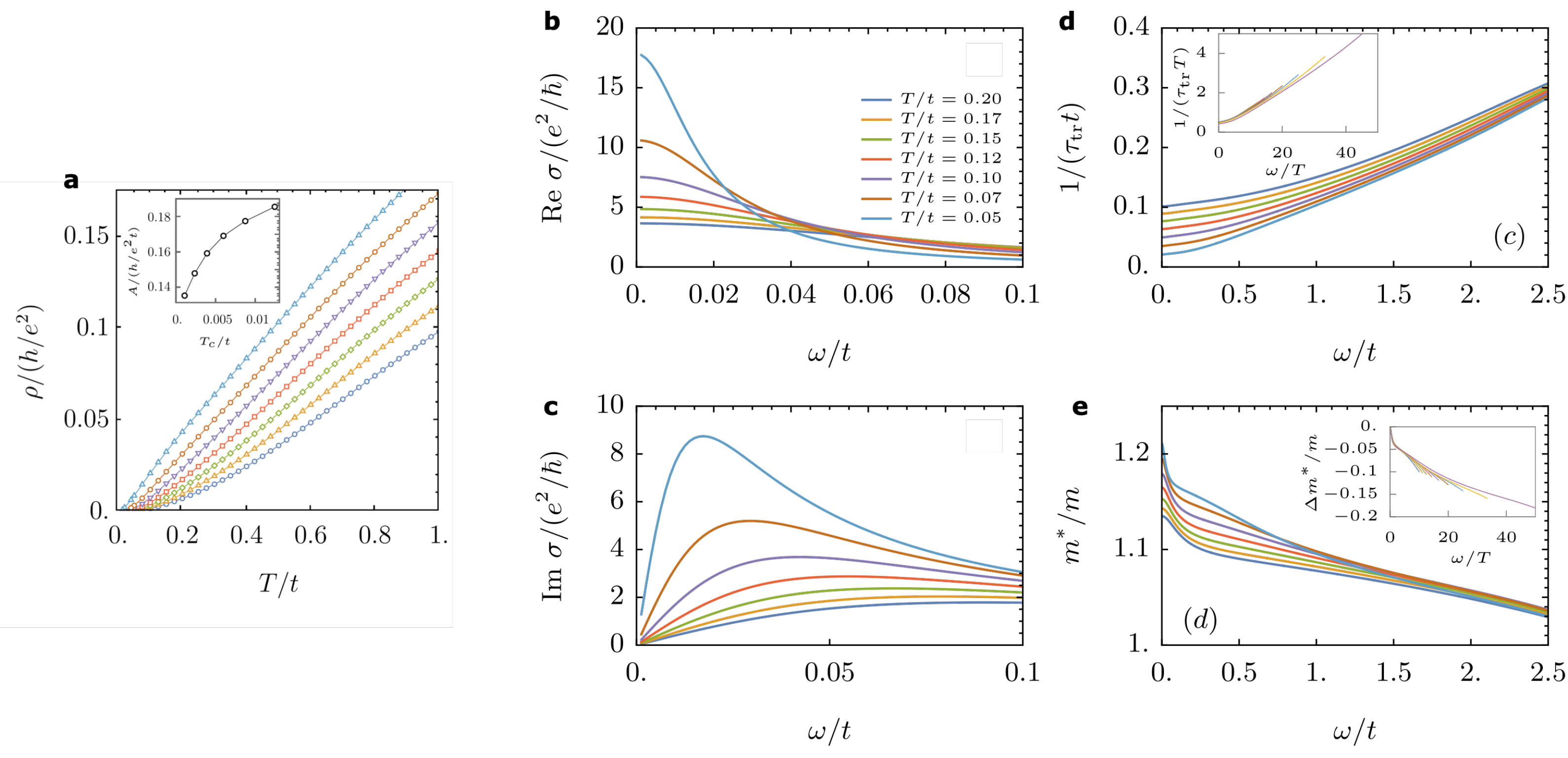}
\caption{(a) Normal state resistivity of the $g'$-model for different values of the renormalized boson mass $M(T=0)$, ranging from $M/t = 1.3$ (top) to $M/t = 0$ (bottom). (b,c) Real and imaginary parts of the optical conductivity in the normal state at the QCP for various temperatures. Panels (d) and (e) show the optical scattering rate $1/\tau_\text{tr}$ and effective mass $m^*$ obtained from $\sigma$ using \eqref{eq:gen_drude}. Inset of panel (c) is a scaling plot with $1/\tau_\text{tr}$ and $\omega$ scaled by $T$; inset of panel (d) is a scaling plot showing $\Delta m^*/m$ as a function of $\omega/T$. Figures reproduced from Ref.~\cite{li2024}.}
\label{fig:gp_rho+sig}
\end{figure}

\subsubsection{$g'$-model} The analysis with the the spatially disordered $g'$ interaction is similar to that for the $g$-model. The most significant difference is that the electron and boson self-energies in this case are local in real space, owing to the $\delta$-correlated spatial disorder:
	\begin{align}
	\Sigma(i\omega_m) &= -g'^2 T \sum_{m'}D_\text{loc}(i\omega_m - i\omega_{m'})G_\text{loc}(i\omega_{m'}), \label{eq:eliash_sig3}
	\\
	\Pi(i\nu_m) &=  -2 g'^2 T \sum_{m'}G_\text{loc}(i\omega_{m'} + i \nu_m) G_\text{loc}(i\omega_{m'}),
    \label{eq:eliash_pi3}
	\end{align}
where
    \be
    G_\text{loc}(i\omega_m) = \int_{\bfk} \frac{1}{i\omega_m - (\epsilon_{\bfk} - \mu) - \Sigma(i\omega_m)}, \quad 
    D_\text{loc}(i\nu_m) = \int_{\bfq} \frac{1}{\nu_m^2 + \omega_{\bfq}^2 - \Pi(i\nu_m)},
    \label{eq:loc_prop}
    \ee
and $\epsilon_{\bfk}$, $\omega_{\bfq}$ are the lattice regularized electron and boson dispersions, respectively. 

Tuning the boson to criticality, a low-energy form of the solution to Eqs.~\eqref{eq:eliash_sig3} and \eqref{eq:eliash_pi3} can be obtained for a quadratic electronic band \cite{esterlis2021,patel2023,li2024}:
	\be
	\Pi(i\nu_m) = -2g'^2 \pi \mathcal N^2 |\nu_m|, \quad \Sigma(i\omega_m) \sim -i g'^2 \mathcal N \omega_m \log\left(\frac{1}{|\omega_m|}\right),
    \label{eq:mfl}
	\ee
where numerical prefactors have been dropped in the expression for $\Sigma$. These self-energies yield a diffusive dynamics for the boson  $D^{-1}(q,i\nu_m) \sim |\nu_m| + \mathcal D q^2$, with $\mathcal D$ the diffusion coefficient, and \textit{marginal Fermi-liquid} electronic behavior \cite{varma1989}. This implies, in particular, an effective mass divergence (and corresponding specific heat anomaly) of the form $m^*/m \sim \log 1/T$ \cite{esterlis2021}.

The marginal Fermi-liquid behavior \eqref{eq:mfl} in the quantum critical fan leads to transport properties of the $g'$-model that are of significant interest in the context of strange metal phenomenology. Unlike the $g$-model, the $g'$-model has, even in the absence of umklapp scattering, a finite dc conductivity due to the fact that the disordered coupling breaks translation invariance \cite{guo2022,patel2023,li2024}\footnote{The scattering to all momenta by the $g'$ coupling also implies that the qualitative phenomenology should also be largely insensitive as to whether the order parameter has $\mathbf{Q} \neq 0$ (like the SDW or CDW).}. Because the self-energy \eqref{eq:eliash_sig3} in the $g'$-model is local in space, vertex corrections for the conductivity vanish \cite{patel2023,khurana1990} and the conductivity is determined by the dressed particle-hole bubble. The frequency-dependence of the single-electron scattering rate $-\im\Sigma_R \sim |\omega|$ (obtained from analytic continuation of \eqref{eq:mfl}) then immediately leads to a dc resistivity $\rho \sim g'^2T$, up to logarithmic corrections, reproducing the hallmark $T$-linear resistivity of strange metals. The resistivity of the $g'$-model obtained from a full numerical solution of \eqref{eq:eliash_sig3} and \eqref{eq:eliash_pi3} discretized on a 2d square lattice is shown in \textbf{Figure \ref{fig:gp_rho+sig}a} (for details see Ref.~\cite{li2024}). Away from the QCP, the resistivity is linear at elevated $T$ inside the quantum critical fan, eventually crossing over to a more conventional $T^2$ behavior in the Fermi-liquid regime at lower temperatures. 

With regard to the quantum-critical behavior, an important and outstanding question concerns the $\omega/T$ scaling of the optical conductivity within the quantum critical fan \cite{Sachdev2000}; a problem that has received significant attention in the cuprates in particular \cite{michon2023,varma2020}. In the $g'$-model the simplest version of such scaling is spoiled by logarithmic corrections, arising from the $T \to 0$ logarithmic divergence of the effective mass, as described above. Nevertheless, following an analysis proposed in \cite{michon2023}, it can be shown that a version of $\omega/T$ still approximately holds. The analysis proceeds through a ``generalized Drude" parametrization of the optical conductivity in the form
    \be
    \sigma(\omega) = i\frac{Ke^2/2}{\omega m^*(\omega)/m + i / \tau_\text{tr}(\omega)}.
    \label{eq:gen_drude}
    \ee
This relation defines the frequency-dependent (or ``optical") scattering rate $1/\tau_\text{tr}(\omega)$ and mass enhancement parameter $m^*(\omega)/m$, which can be obtained from the real and imaginary parts of $\sigma(\omega)$ once the optical weight $K$ is known. The weight $K$ is equal to the average electronic kinetic energy and can thus be separately determined. 

The optical conductivity, optical scattering rate, and optical mass obtained from the numerical solution of the $g'$-model on a 2d square lattice are shown in  \textbf{Figure \ref{fig:gp_rho+sig}b-e} \cite{li2024}. The optical scattering rate is linear in frequency down to frequency $\omega \sim T$, and the inset of \textbf{Figure \ref{fig:gp_rho+sig}c} demonstrates a reasonable $\omega/T$ scaling collapse. To investigate $\omega/T$ scaling of the optical mass, the combination $\Delta m^*/m = m^*(\omega)/m - m^*(0)/m$ is used in order to subtract the logarithmically divergent contribution that violates scaling. This quantity is shown in \textbf{Figure \ref{fig:gp_rho+sig}e}. As may be seen from the inset of the figure, to the extent $\omega/T$ scaling is observed it is only over a rather narrow range of $T$, likely being spoiled by additional logarithmic corrections. Taken together, these properties of the optical conductivity of the $g'$-model demonstrate an encouraging level of qualitative agreement to observations in the cuprates, as analyzed in \cite{michon2023}.

In addition to the optical conductivity, measurements of the dynamic charge susceptibility in the strange-metal regime of the cuprates, obtained via momentum-resolved inelastic electron scattering, also exhibit an approximate power-law scaling behavior in the density response \cite{mitrano2018,husain2019}, suggestive of some form of criticality. Whether similar behavior can be reproduced within the YSYK framework is a subject of ongoing investigation \cite{wang2023_density,klein_toappear}. 

Note that the analysis presented here applies only to strictly two-dimensional systems. For $d=3$, or more generally, for any dimension $d>2$, one finds conventional Fermi liquid behavior in the $g'$-model. This result  follows from the  fact that the local propagator $D_{\rm loc}(i\nu_m)$ of Eq.~\eqref{eq:loc_prop} is less singular in this case.

\subsection{Significance and Takeaways} 
The generalization of the YSYK model to higher dimensions yields the finite-dimensional Eliashberg equations as a large-$N$ solution. In higher dimensions, crucial effects associated with the existence of the Fermi surface are incorporated beyond the quantum dot model, and many existing results regarding quantum critical Fermi surfaces are unified within a systematic large-$N$ expansion. Beyond providing a theoretical construct to rationalize known results, new insights have been obtained by incorporating the  spatially inhomogeneous Yukawa coupling $g'(x)$ into the YSYK model, which serves as a simple way to incorporate ``random mass" disorder in the position of the QCP. This form of coupling leads to a marginal Fermi liquid state, where the dc and ac responses are strikingly similar to that observed in strange metal superconductors, suggesting that this form of disorder may play a central role in determining the salient properties of these systems. 

\section{Eliashberg theory and holography}
\label{sec:holography}

One of the challenges in the theory of strongly-correlated materials is the sparseness of theoretical tools that allow for a description of systems without quasi-particle excitations. The  holographic correspondence \cite{Maldacena1998}, i.e. the duality between quantum field theories and gravity theories in asymptotically Anti-de-Sitter (AdS) spaces in one higher dimension, was therefore a welcomed addition to the arsenal of theory concepts that allow for a more concrete description, in particular of quantum critical systems. The scaling invariance, typical for quantum critical points, can be understood either as an effect of the diverging correlation length or correlation time. Alternatively, it can be rationalized as a geometric property of AdS spaces. Within the holographic correspondence, quantum mechanical operators and external fields that couple to them, i.e. external sources,  are mapped into gravity fields.  The generating functional of the physical theory is then given by the holographic Euclidean action $S$ as
	\begin{equation}
		\left\langle e^{-\int d^{d+1}x J {O}}\right\rangle =\left.e^{-S\left[\psi\right]}\right|_{\psi\left(\zeta\to0\right)=J}.
		\label{GKPW}
	\end{equation}
	Here $O$ is an operator of the ($d+1$)-dimensional field theory with source $J$, $\psi$ is the dual field, and $\zeta$ is the coordinate in the extra dimension, with $\zeta=0$ identifying the boundary of the AdS space \cite{Witten1998,Gubser1998}. The evolution along the extra direction is often interpreted as encoding the RG flow of the dual quantum system in a way that small values of $\zeta$ probe the system at high energies, while  large $\zeta$  corresponds to the low energy, universal regime \cite{DeHaro2001,Skenderis_2002}. 	Phrasing a problem in this geometric language allows for the analysis of phenomena as diverse as thermoelectric transport without quasi-particles \cite{Donos2014,PhysRevB.76.144502}, the quark-gluon plasma \cite{casalderrey-solana_liu_mateos_rajagopal_wiedemann_2014}, and strongly coupled hydrodynamics \cite{PhysRevLett.94.111601,PhysRevLett.87.081601}. Moreover, holography provides simple built-in mechanisms for spontaneous symmetry breaking that have been used to gain insight into, e.g., high $T_\text{c}$ superconductors \cite{Gubser2008,Hartnoll2008,Hartnoll_2008HEP} or charge-density
	waves \cite{Donos2011,PhysRevB.96.195128,1603.03029}. Despite these successful applications, it is fair to say that the formalism and physical interpretation of the holographic correspondence is not equally natural to all condensed matter physicists. To have an explicit model that allows for a step-by-step derivation of the holographic correspondence may prove useful to gain intuition. In addition, the holographic formulation offers in some situations an easier and more direct formulation of superconductivity in quantum-critical metals.  In what follows we will show that one can  explicitly derive the correspondence of Eq.\eqref{GKPW} for the $0+1$-dimensional quantum dot problem discussed in section~\ref{sec:quantum_dot}. The analysis of this section closely follows Ref.~\cite{Inkof2021}, while extensions to higher spatial dimensions are discussed in Ref.~\cite{2209.00474}.

    \subsection{Holographic superconductivity}
We start with a very brief summary of  holographic superconductivity, a  concept that was introduced to formulate 
a gravity dual to a superconducting phase transition\cite{Gubser2008,Hartnoll2008,Hartnoll_2008HEP}. Following the usual philosophy of gravity theory, we consider the action 
\begin{equation}
S=S_{{\rm EH}}+S_{{\rm gauge}}+S_{{\rm matter}},
\end{equation}
where $S_{{\rm EH}}=\int d^{D}x\sqrt{-g}\left(R-\Lambda\right)$ is
the Einstein-Hilbert action with metric tensor $g_{\mu\nu}$ and determinant
$g$, Ricci scalar $R$, and cosmological constant $\Lambda$. $S_{{\rm gauge}}$
describes the electromagnetic field and $S_{{\rm matter}}$ the matter
that exists in this space-time. The Euler-Lagrange equation of this
problem with respect to $g_{\mu\nu}$ then yields the Einstein equations.
For $\Lambda=0$ the homogeneous solution of $\frac{\delta S_{{\rm EH}}}{\delta g_{\mu\nu}}=0$
is asymptotically flat, while it has positive (negative) curvature
for a positive (negative) cosmological constant. $\Lambda>0$ then
corresponds to de Sitter space dS$_{D}$, while $\Lambda<0$ is anti-de
Sitter AdS$_{D}$, with space-time dimension $D$. 

For our analysis we consider a gravitational theory $S_{{\rm EH}}$
that leads asymptotically to a $D=d+2$ dimensional anti-de Sitter
space. We will work in imaginary time, i.e.
we write $\tau=-it$. Our main interest is AdS$_2$, which is equivalent to the two-dimensional hyperbolic space $H_2$, parametrized via $X_1^2+X_2^2-X_3^2=-1$ with the three-dimensional Cartesian coordinates $(X_1,X_2,X_3)$. A convenient description is achieved in terms of Poincare coordinates that correspond to $X_1=\tau/\zeta$ and $X_{2,3}=\frac{\zeta}{2}\left( 1+\zeta^{-2}(\tau^2\mp 1)\right)$, with imaginary time $\tau$ 
as well as the radial coordinate $\zeta\in\left[0,\infty\right]$. This immediately yields the induced differential distance $ds^{2}=\zeta^{-2}\left(d\tau^{2}+d\zeta^{2}\right)$.
Similarly one can  use for higher-dimensional systems Poincare coordinates in AdS$_D$:$x^{\mu}=\left(\tau,\zeta,\boldsymbol{x}\right)$ with the $d$-dimensional spatial part $\boldsymbol{x}$, and distance
$ds^{2}  =  g_{\mu\nu}dx^{\mu}dx^{\nu}=\frac{1}{\zeta^{2}}\left(d\tau^{2}+d\zeta^{2}+d\boldsymbol{x}^2\right)$.
At fixed $\zeta$, the metric describes a flat $d+1$-dimensional space time; only the ``new'' coordinate $\zeta$ is responsible for the curvature. One also observes that the distance between two points $x^{\mu}$ and $x'^{\mu}$ is very large for small $\zeta$. Hence, a geodesic between two points near the boundary $\zeta=0$ will always go into the bulk (where $\zeta$ is no longer small) and then turn back.

The part $S_{{\rm gauge}}$ is given by the Maxwell action in curved space
and is not of primary interest for the present discussion. Finally, the matter degrees of freedom 
in our problem are described by the action of a charged, i.e. complex,
scalar field in AdS$_{D}$:
\begin{equation}
S_{{\rm matter}}=\int d^{D}x\sqrt{g}\left[D_{\mu}\psi^{*}D^{\mu}\psi+m^2\psi^{*}\psi+u\left(\psi^{*}\psi\right)^{2}\right],\label{eq:holSC}
\end{equation}
where we included leading gradient terms and nonlinear terms in the
expansion up to quartic order, and $D_{\mu}=\partial_{\mu}-ie^\ast A_{\mu}$
with $e^\ast$ the charge of the scalar. Eq.~\eqref{eq:holSC} is the Ginzburg-Landau theory of a superconductor
in curved space and with one extra dimension. The  theory of Eq.~\eqref{eq:holSC} becomes unstable towards condensation of $\psi$ once  $m^2$ is below a threshold value. In flat space the instability at the Gaussian level occurs once $m^2<0$.  As shown by Breitenlohner and
Freedman \cite{Breitenlohner1982},  condensation  occurs in a space of negative curvature only once  $m^2<m^2_{\rm BF}<0$ with $m^2_{\rm BF}=-\frac{1}{4}\left( d+1\right)^2$. 

The questions related to Eq.~\eqref{eq:holSC} that  we wish
to discuss are: i) What is the physical essence of this extra dimension $\zeta$
within a concrete many-body calculation? ii) Why do we even need this
extra coordinate? iii) What was wrong with the good-old-fashioned
Ginzburg-Landau description? To give the answers right away: i) The
extra dimension describes the internal dynamics of the Cooper pair
as a composite object that consists of two fermions. ii) The physical
origin of the geometric formulation in the curved space is the unique scale invariance of a quantum-critical normal state that is highly non-local in time. iii) The crucial internal
dynamics of the Cooper pair must be included for superconductivity in scale-invariant
quantum critical systems, forcing us to generalize the known Ginzburg-Landau
formalism. 

\subsection{SYK-superconductor and holography at $T=0$}    
   
We start with  a simple physical
motivation that illustrates the close relationship between the Eliashberg
formalism and behavior in curved space. To this end, we demonstrate that the linearized gap equation can be recast in the form of a wave equation in Anti-de Sitter space.  Let us start from the
linearized gap equation \eqref{eq:universal_phi}
that follows from the gap equation Eq.~\eqref{eq:gapeq} if one uses the quantum-critical power law results of Eq.~\eqref{eq:syknfl}. 
We introduce a lower cut off $\sim T$
at small but non-zero temperatures, along with an upper cut off $\Lambda$
above which these power laws cease to be correct. In the limit of the exponent $\Delta$ near $1/4$, 
one can easily rewrite the resulting integral equation as a differential
equation. We split the integral over over $\epsilon'$
into $\left|\epsilon'\right|\in[T,\epsilon]$ and $\left|\epsilon'\right|\in[\epsilon,\Lambda]$
and approximates $\left|\epsilon-\epsilon'\right|^{\gamma}\approx\max\left(\left|\epsilon\right|^{\gamma},\left|\epsilon'\right|^{\gamma}\right)$, where $\gamma=1-4\Delta \ll 1$. 
Then it follows that
\begin{equation}
\frac{d}{d\epsilon}\epsilon^{1-\gamma}\frac{d}{d\epsilon}\epsilon^{\gamma}\Phi\left(i\epsilon\right)=-\frac{\gamma \kappa}{\pi}\frac{\Phi\left(i\epsilon\right)}{\epsilon},
\label{eq:dgl01}
\end{equation}
along with appropriate boundary conditions~\cite{hauck2020}. Here $\kappa = (1-\alpha)/C_\Delta$; see Eq.~\eqref{eq:universal_phi}. If we now introduce
the new variable $\zeta=1/\epsilon$ and define the function 
$\psi\left(\zeta\right)=\zeta^{\frac{1-\gamma}{2}}\Phi\left(i/\zeta\right)$,
Eq.~\eqref{eq:dgl01} becomes
\begin{equation}
    -\partial_{\zeta}^{2}\psi+\frac{m^{2}}{\zeta^{2}}\psi=0,
\end{equation}
with mass $m^{2}=f(\gamma,\kappa)$ a function of parameter $\kappa$ and the exponent $\gamma$.
This is the Klein-Gordon equation of a scalar field $\psi\left(\zeta,\tau\right)$
in AdS$_{2}$ that is independent of the time $\tau$. The onset of pairing can be shown to coincide with $m^{2}=m^{2}_{\rm BF}$, where $m^{2}_{\rm BF}=-1/4$ is the Breitenlohner and
Freedman bound \cite{Breitenlohner1982} of a system in ${\rm AdS}_2$. Hence, the
critical normal state that is responsible for the power-law structure
of the theory gives rise to a geometric structure
in curved space and offers a first hint on how to construct the scalar
field of the holographic theory. On the other hand, this approach
is only useful for time-independent solutions, where we will see that
$\tau$ stands for the absolute time of the superconducting correlations
while $\epsilon=\zeta^{-1}$ is related to the frequency of the Fourier transform of the relative time. 

Next we extend the analysis and allow for an arbitrary dependence
on the absolute and relative time of pairing correlations. We recall
that the derivation of the large-$N$ equations within the SYK-formalism
is based on introducing bilocal fields that, at the saddle point,
amount to propagators and self energies, i.e. $Z=\int DG\left(\tau,\tau'\right)D\Sigma\left(\tau,\tau'\right)e^{-NS}$.
Hence, the fluctuating variables depend on two time coordinates that
we can always rewrite in terms of $(\tau+\tau')/2$ and $\tau-\tau'$.
In models with a superconducting solution the approach is then generalized
and includes, in addition to the bosonic propagators and self energies,
also anomalous Gor'kov propagators and self energies as fluctuating variables, i.e.
\begin{equation}
Z=\int DG\cdots DFD\Phi D\bar{F}D\bar{\Phi}e^{-NS},
\end{equation}
where the conditions $F(\tau,\tau')=c_{\uparrow}(\tau)c_{\downarrow}(\tau')$
and $\bar{F}(\tau,\tau')=c_{\downarrow}^{\dagger}(\tau)c_{\uparrow}^{\dagger}(\tau')$
are enforced by $\Phi\left(\tau,\tau'\right)$ and $\bar{\Phi}\left(\tau,\tau'\right)$,
respectively. The action $S$ is rather lengthy and has
been discussed in the literature~\cite{inkof2022quantum}. If we wish to study small superconducting
fluctuations of a quantum-critical normal state, we can expand the
action up to quadratic order in the pairing fields, i.e. $S\approx S^{\left(0\right)}+S^{\left({\rm sc}\right)}$
with $S^{\left(0\right)}$ the normal state action and 
\begin{equation}
S^{\left({\rm sc}\right)}=\int_{\omega,\epsilon}\frac{\bar{F}\left(\omega,\epsilon\right)F\left(\omega,\epsilon\right)}{\Pi_{{\rm n.s.}}\left(\omega,\epsilon\right)}-\frac{g^{2}(1-\alpha)}{2}\int_{\omega,\epsilon,\epsilon'}\bar{F}\left(\omega,\epsilon\right)D_{{\rm n.s.}}\left(\epsilon-\epsilon'\right)F\left(\omega,\epsilon'\right).\label{eq:S_gauss_SC}
\end{equation}
Here $\omega$ is the Fourier transform of the averaged time
$(\tau+\tau')/2$ while $\epsilon$ is the Fourier transform of the
relative time $\tau-\tau'$. $\Pi_{{\rm n.s.}}\left(\omega,\epsilon\right)=G_{{\rm n.s.}}\left(\frac{\omega}{2}-\epsilon\right)G_{{\rm n.s.}}\left(\frac{\omega}{2}+\epsilon\right)$
is the particle-particle correlation function, expressed in terms
of the normal state fermionic propagators, while $D_{{\rm n.s.}}(\epsilon)\propto|\epsilon|^{-\gamma}$
is the bosonic propagator. This is a Gaussian action for a field that
lives in a two-dimensional space, even though formally we are considering
a $(0+1)-$dimensional problem with zero space dimensions $d=0$ and
one time axis. The situation becomes even clearer if one writes 
\begin{equation}
F\left(\frac{\tau_{1}+\tau_{2}}{2},\epsilon\right)=\left|\epsilon\right|^{\frac{\gamma-1}{2}}\int_{\Gamma}\psi\left(\tau,\zeta\right)dl, \label{eq:map}
\end{equation}
where the integration is over the contour $\Gamma$ determined by
the condition
$\left|\epsilon\right|^{-2}=
\left(\tau-\frac{\tau_{1}+\tau_{2}}{2}\right)^{2}+\zeta^{2}$,
which  describes a geodesics in AdS$_{2}$
and the non-local relationship Eq.~\eqref{eq:map} is referred to as
a Radon transformation~\cite{Das2018}. We will now see that Eq.~\eqref{eq:map} is
indeed the explicit holographic map from the quantum-many body theory
in $0+1$ dimensions to the holographic superconductor in $D=2$ dimensions,
i.e. in one extra space-time dimension. Inserting Eq.~\eqref{eq:map}
into the action Eq.~\eqref{eq:S_gauss_SC}, and using properties of
the Radon transformation discussed in Ref.~\cite{Das2018}, yields within a
gradient expansion 
\begin{equation}
S^{\left({\rm sc}\right)}=\int d\tau d\zeta\left(\frac{m^{2}}{\zeta^{2}}\left|\psi\right|^{2}+\left|\partial_{\tau}\psi\right|^{2}+\left|\partial_{\zeta}\psi\right|^{2}\right).
\end{equation}
This is precisely the action of a holographic superconductor in Poincaré
coordinates on AdS$_{2}$ at the Gaussian level (c.f. Eq.~\eqref{eq:holSC}).
Hence, we have established the direct link between the critical Eliashberg theory discussed in section~\ref{sec:quantum_dot} and  holographic superconductivity. Both are different formulations of the the same theory. 

\subsection{Holographic mapping at finite $T$ and chemical potential}

\begin{figure}[t]
\centering {\includegraphics[width=3.3in]{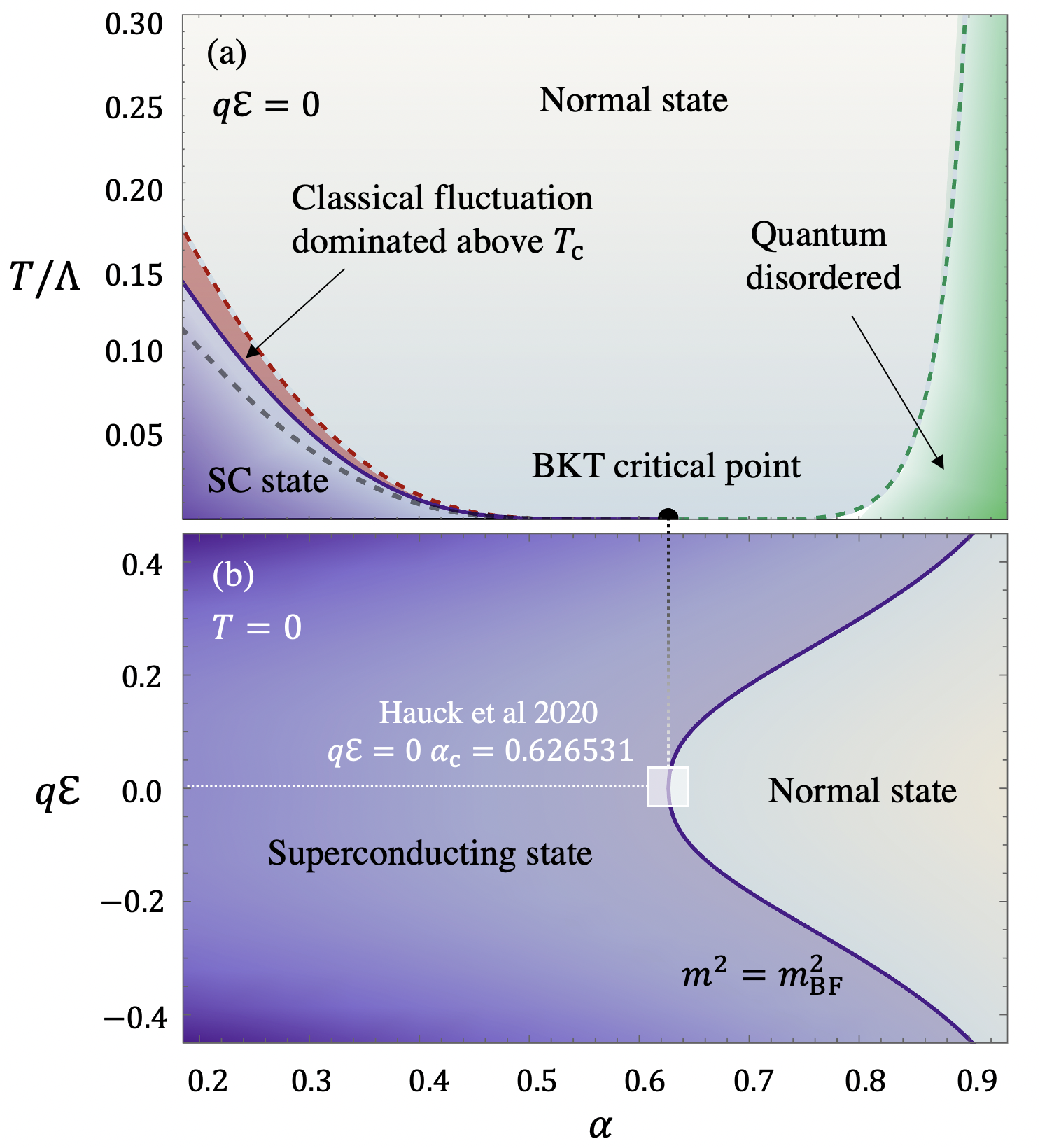}} \caption{ (a) Phase diagram of the Yukawa-SYK model spanned
by temperature in units of the upper cutoff $T/\Lambda$ and pair
breaking parameter $\alpha$ at charge neutrality $q\mathcal{E}=0$. At $\alpha_{\text{c}}=0.623\pm0.03$
the critical temperature vanishes in a BKT fashion \cite{hauck2020}.
For $\alpha<\alpha_{\text{c}}$ the system becomes superconducting
as temperature is lowered. The blue area represents the symmetry-broken
phase. The red one is the classical fluctuation dominated regime where
the pairing susceptibility diverges following a Curie-Weiss law $\chi\propto(T-T_{\text{c}})^{-1}$. The quantum critical
phase transition is signaled by the derivatives of the pairing susceptibility,
that have been used to estimate the crossover from the quantum critical
to quantum disordered regime on the $\alpha>\alpha_{\text{c}}$ side.
Figure re-adapted from Ref.\cite{Inkof2021}. (b) Phase diagram
of the Yukawa-SYK model spanned by $q\mathcal{E}$ and pair breaking
parameter $\alpha$. $q$ is the fermion charge and ${\cal E}$ the
spectral asymmetry. 
Once the system reaches the Breitenlohner-Freedman bound 
the superconducting phase sets in.}
\label{Phase_Diagram_density} 
\end{figure}

The holographic mapping can also be extended to finite temperatures.
To this end, one exploits an important re-parametrization invariance
of the low-energy theory \cite{chowdhury2022} that allows one to obtain the finite temperature
propagators $G$ and $D$ of the normal state from the $T=0$ solutions
$G_{T=0}\left(\tau_{0}\right)$ and $D_{T=0}\left(\tau_{0}\right)$
via re-parametrization of the time axis $\tau\rightarrow\tau_{0}=f\left(\tau\right)$. In our problem $f\left(\tau\right)=\frac{1}{\pi T}\tan\left(\pi\tau T\right)$
maps the interval $-\frac{1}{2 T}<\tau<\frac{1}{2 T}$ onto $-\infty<\tau_{0}<\infty$.
It then follows that 
\begin{eqnarray}
G\left(\tau,\tau'\right) & = & f'\left(\tau\right)^{\Delta}G_{T=0}\left(f\left(\tau\right)-f\left(\tau'\right)\right)f'\left(\tau'\right)^{\Delta},\nonumber \\
D\left(\tau,\tau'\right) & = & f'\left(\tau\right)^{1-2\Delta}D_{T=0}\left(f\left(\tau\right)-f\left(\tau'\right)\right)f'\left(\tau'\right)^{1-2\Delta}.
\end{eqnarray}
 Hence, the finite temperature problem leads to an effective action,
identical to the one we just analyzed at $T=0$, but now  in terms of the
field 
\begin{equation}
    {\cal F}\left(\tau_{0},\tau_{0}'\right)=\left(1+\left(\pi T\right)^{2}\tau_{0}^{2}\right)^{-\Delta}F\left(f^{-1}\left(\tau_{0}\right),f^{-1}\left(\tau_{0}'\right)\right)\left(1+\left(\pi T\right)^{2}\tau_{0}'^{2}\right)^{-\Delta}.
\end{equation}
We can now  establish a connection that is analogous
to Eq.~\eqref{eq:map} only now in terms of the new function ${\cal F}$,
instead of $F$, and with transformed time. At the same time it is
possible to transform the resulting holographic action to coordinates
$\left(\tau,\zeta\right)$ that agree at the boundary with the original
imaginary time variables \cite{Sachdev2019}:
\begin{eqnarray}
\tau_{0} & = & 2\zeta_{0}\frac{\left(1-\zeta^{2}/\zeta_{T}^{2}\right)^{1/2}\sin\left(\tau/\zeta_{T}\right)}{1+\left(1-\zeta^{2}/\zeta_{T}^{2}\right)^{1/2}\cos\left(\tau/\zeta_{T}\right)},\nonumber \\
\zeta_{0} & = & \frac{2\zeta}{1+\left(1-\zeta^{2}/\zeta_{T}^{2}\right)^{1/2}\cos\left(\tau/\zeta_{T}\right)},\label{eq:finiteTtransf}
\end{eqnarray}
where $\zeta_{T}=1/\left(2\pi T\right)$. This guarantees that the theory obeys the proper boundary conditions at finite temperature. In these coordinates the
finite-$T$ action of the SYK superconductor becomes once again a
holographic superconductor according to Eq.~\eqref{eq:holSC}. The crucial difference
is that at finite $T$ the metric of the problem is now given by 
\begin{equation}
ds^{2}=\frac{l^{2}}{\zeta^{2}}\left(\left(1-\zeta^{2}/\zeta_{T}^{2}\right)d\tau^{2}+\frac{d^{2}\zeta}{1-\zeta^{2}/\zeta_{T}^{2}}\right).
\end{equation}
This is the metric of AdS$_{2}$ with black hole and horizon $\zeta_{T}$ determined by the corresponding Hawking temperature.
Hence, one can derive from a Hamiltonian of interacting electrons
and bosons that the natural description
of the finite-temperature pairing fluctuations in a quantum critical regime is governed by a metric
with a black hole. Fluctuations  described by the variable $\zeta$ are cut off by
the black hole horizon for energies below $T$. 

The correspondence can not only be extended to finite temperatures
but also to finite chemical potential $\mu$, i.e. away from the particle-hole
symmetric point. At $T=0$ but finite $\mu$  a generalization of Luttinger's theorem \cite{Luttinger1961}
for systems with anomalies of the power law form was established \cite{georges2001,Gu2020}:
\begin{equation}
n=\frac{1}{N}\sum_{i=1}^{N}\left\langle c_{i\sigma}^{\dagger}c_{i\sigma}\right\rangle =\frac{1}{2}-\frac{\theta}{\pi}-\left(\frac{1}{2}-\Delta\right)\frac{\sin\left(2\theta\right)}{\sin\left(2\pi\Delta\right)}.\label{eq:Lutt}
\end{equation}
Here $\Delta$ is still the anomalous fermionic exponent and $\theta\in\left[-\pi\Delta,\pi\Delta\right]$ is a measure
of the deviation from half filling. We will express it as $\tan\theta=\tan\left(\pi\Delta\right)\tanh\left(\pi e{\cal E}\right)$. The electron charge is $e$ and the spectral asymmetry parameter ${\cal E}$ was discussed in Ref.\cite{sachdev2015,Sachdev2019}
where the relation $2\pi{\cal E}=\frac{\partial S_{0}}{\partial n}$
to the density dependence of the zero-point entropy was established.

While the expression for the bosonic propagator is unaffected by the
chemical potential, the fermionic one is now given by 
\[
G\left(\tau,\tau'\right)=\frac{g\left(\tau'\right)}{g\left(\tau\right)}f'\left(\tau\right)^{\Delta}G_{T=\mu=0}\left(f\left(\tau\right)-f\left(\tau'\right)\right)f'\left(\tau'\right)^{\Delta},
\]
with $g\left(\tau\right)=e^{-2\pi e{\cal E}T\tau}$, i.e. the low-energy
saddle point equations are invariant under re-parametrization $\tau\rightarrow\tau_{0}=f\left(\tau\right)$
and $U(1)$ transformations. Notice, $g\left(\tau\right)$
becomes a genuine $U(1)$ transformation once we return
to real times $\tau=-it$. This allows one to demonstrate that the particle-particle
fluctuation spectrum can be expressed in terms of the one at $\mu=0$
via 
\begin{equation}
\Pi\left(\epsilon_{n},\omega_{m}\right)=\Pi_{\mu=0}\left(\epsilon_{n},\omega_{m}-i4\pi e{\cal E}T\right).
\end{equation}
In Ref.~\cite{Faulkner2011} it was shown that the propagator of
a charge $e^{*}$ scalar particle within AdS$_{2}$ in an boundary
electric field $\cal{E}$, described within the gauge $A_{\zeta}=0$
by 
\begin{equation}
A_{t}=\frac{{\cal E}}{\zeta}\left(1-\frac{\zeta}{\zeta_T} \right),
\end{equation}
is determined by a shift of the propagator as $\omega\rightarrow\omega+2\pi e^{*}{\cal E}T$
relative to the neutral case. Hence, we can conclude that the boundary
electric field ${\cal E}$ is determined by the deviation of
the SYK particle density $n$ from half-filling with effective charge
$e^{*}=2e$. Including a chemical potential in
the original many-body theory that lives on the boundary, i.e. in the actual quantum-field theory, corresponds
to including a local $U(1)$ vector potential with an associated electrical
field that is determined by the spectral asymmetry parameter ${\cal E}$. The spectral asymmetry parameter ${\cal E}$, i.e. the deviation in the carrier concentration from its particle-hole symmetric half-filled state, changes the value of the Breitenlohner Freedman bound and hence allows determining the onset of pairing as a function of carrier concentration. The resulting phase diagram at half filling is shown in \textbf{Figure \ref{Phase_Diagram_density}a}. The density dependence of the critical pair-breaking parameter $\alpha_c$ is shown in \textbf{Figure \ref{Phase_Diagram_density}b}.

\subsection{Significance and Takeaways} 
In this section, we have demonstrated a direct and explicit connection between quantum-critical Eliashberg theory and the theory of holographic superconductivity. At low energies, the two theories are identical. 
Within the Eliashberg formalism, the critical normal state gives rise to a scale-invariant, non-Fermi liquid bosonic and fermionic spectrum that changes the nature of the pairing instability and gives rise to power law dependencies of the anomalous self-energy near the onset of superconductivity. From the holographic perspective,  the same scale-invariant quantum critical normal state endows the theory with a non-trivial geometry, while  superconducting fluctuations act as matter fields in this curved space. Here we demonstrated this equivalence at the Gaussian level. One can extend the analysis to include, e.g., a quartic self-interaction of the scalar field \cite{inkof2022quantum}, exactly as given in Eq.~\eqref{eq:holSC}. The specific holographic correspondence Eq.~\eqref{eq:map} is non-local, i.e. a point in the field theory corresponds to a curve (a geodesic) in anti-de Sitter space. It is nevertheless sensible to interpret the extra dimension of the problem as describing the internal dynamics of the composite Cooper pair, determined by the relative time $\tau-\tau'$ of the pairing correlation function.

The derivation of the holographic correspondence  bears a certain analogy to Gor'kov's 1959 derivation of the Ginzburg-Landau theory of superconductivity from BCS theory \cite{gorkov1959}. In that case, Gor'kov identified the scalar order parameter describing the relevant symmetries of the problem with the anomalous propagator, 
such that  $\psi\left(\boldsymbol{r},\tau\right)\sim F\left(\boldsymbol{r},\tau;\boldsymbol{r},\tau\right)$.
Moreover, he demonstrated that the effective charge of this scalar field is twice the electron charge, $( e^\ast = 2e )$, an insight absent in the original Ginzburg-Landau formulation~\cite{ginzburg2009theory}. His derivation also established direct connections between the parameters of the phenomenological theory—such as the superconducting coherence length, penetration depth, and heat capacity jump—and the underlying microscopic model.  
The derivation of the holographic action from quantum-critical Eliashberg theory presented here follows a similar logic. It provides a microscopic foundation for the meaning of the holographic scalar field and offers explicit expressions for the parameters of the theory in terms of the original microscopic model. 
Beyond conceptual insight, this bridge may have practical significance. Just as the Ginzburg-Landau model, despite being formally derivable from BCS theory, often proves far more convenient in applications, the holographic approach may offer an efficient framework for studying spatially inhomogeneous or out-of-equilibrium critical superconductors—regimes where the direct application of Eliashberg theory becomes impractical. Even when the weak-coupling assumption employed by Gor'kov breaks down, the Ginzburg-Landau model remains a robust and insightful tool. Similarly, the holographic approach may provide a powerful alternative in contexts where Eliashberg theory reaches its limits.  

\section{Conclusions and outlook}
\label{sec:conclusions}

Questions concerning the interplay of quantum criticality, strange metallicity, and superconductivity have long been at the forefront of research on correlated electron systems. The family of YSYK -- or ``quantum critical Eliashberg" -- models described in this review offer a controlled large-$N$ framework within which to address some of these questions. Although the models are not meant to describe the realistic microscopics of any specific system, a number of general lessons and new insights have emerged. We have attempted to summarize the most important conclusions at the end of each section throughout the article. 

On the one hand, many existing results have been unified within a controlled theoretical framework: the Eliashberg equations emerge as the large-$N$ saddle point of a theory in which the Yukawa couplings are random in flavor space, and a systematic procedure by which to compute corrections in terms of bilocal fields has been laid out. Perhaps more importantly, however, entirely new results and insights have been gained. This includes the elucidation of spatially inhomogeneous couplings as a mechanism for reproducing certain key properties of strange metals, including their dc and ac transport. Additionally, the explicit connection  between the YSYK model and holographic superconductivity may serve as a useful new tool. Much as the traditional Landau-Ginzburg approach allows for the analysis of a variety of superconducting effects that are challenging to address from a micropscopic BCS-like approach, the explicit connection to  holographic superconductivity presents a route by which to address novel phenomena in quantum critical superconductors, e.g., properties of collective modes and non-equilibrium effects. 

Further investigation is needed to address these interesting questions, as well as to understand the ultimate limitations of the YSYK approach. Addressing the latter will be most straightforward by comparison with numerically exact methods at finite-$N$ and without flavor-space randomness \cite{patel2024_dqmc}. Although the Eliashberg approximation has been demonstrated to be remarkably accurate in many cases \cite{wang2017,klein2020,xu2020,chubukov2020,chowdhury2020_eliash}, its limitations have also been documented \cite{chubukov2020,zhang2024}. It is of significant interest to incorporate other realistic features of correlated electron systems into the framework; recent steps in this direction include the coupling to dynamical two-level systems modeling metallic glasses \cite{bashan2024_2-level-sys}, and the incorporation of direct Coulomb repulsion \cite{hardy2025}.

%Disclosure
\section*{DISCLOSURE STATEMENT}
The authors are not aware of any affiliations, memberships, funding, or financial holdings that
might be perceived as affecting the objectivity of this review. 

% Acknowledgements
\section*{ACKNOWLEDGMENTS}
We are grateful to Erez Berg, Andrey V. Chubukov, Luca Delacrétaz, Eugene Demler, Haoyu Guo, Sean Hartnoll, Gian-Andrea Inkof, Steven A. Kivelson, Alex Levchenko, Chenyuan Li, Aavishkar Patel, Subir Sachdev, Koenrad Schalm, Yoni Schattner,  Veronika Stangier,  Davide Valentinis, Yuxuan Wang for stimulating discussions.
We would like to thank  Gian-Andrea Inkof for preparing Fig.~\ref{Phase_Diagram_density}. I. E. was supported by the University of Wisconsin–Madison. 
J.S. was supported by the German Research Foundation (DFG) through
CRC TRR 288 “ElastoQMat,” project A07 and the
Simons Foundation Collaboration on New Frontiers in
Superconductivity (Grant SFI-MPS-NFS-00006741-03).
%, and the Gordon and Betty Moore Foundation’s EPiQS Initiative through Grant GBMF4302 while visiting the Geballe Laboratory for Advanced Materials at Stanford University.

\bibliographystyle{ar-style4}
\bibliography{quantum_critical_eliashberg_v2}

\begin{thebibliography}{214}
\expandafter\ifx\csname natexlab\endcsname\relax\def\natexlab#1{#1}\fi

\bibitem{landau1957}
Landau LD. 1957.
\textit{Soviet Physics Jetp-Ussr} 3(6):920--925

\bibitem{agd}
Abrikosov AA, Gorkov LP, Dzyaloshinski IE. 2012.
Methods of quantum field theory in statistical physics.
Courier Corporation

\bibitem{baym2008landau}
Baym G, Pethick C. 2008.
Landau fermi-liquid theory: concepts and applications.
John Wiley \& Sons

\bibitem{polchinski1992}
Polchinski J. 1992.
\textit{arXiv preprint hep-th/9210046}

\bibitem{shankar1994}
Shankar R. 1994.
\textit{Rev. Mod. Phys.} 66(1):129--192

\bibitem{bcsshort}
Bardeen J, Cooper LN, Schrieffer JR. 1957{\natexlab{a}}.
\textit{Phys. Rev.} 106(1):162--164

\bibitem{bcslong}
Bardeen J, Cooper LN, Schrieffer JR. 1957{\natexlab{b}}.
\textit{Phys. Rev.} 108(5):1175--1204

\bibitem{kohn1965}
Kohn W, Luttinger JM. 1965.
\textit{Phys. Rev. Lett.} 15(12):524--526

\bibitem{Cooper1956}
Cooper LN. 1956.
\textit{Physical Review} 104(4):1189

\bibitem{legros2019}
Legros A, Benhabib S, Tabis W, Lalibert{\'{e}} F, Dion M, et~al. 2019.
\textit{Nature Physics} 15(2):142--147

\bibitem{hayes2016}
Hayes IM, McDonald RD, Breznay NP, Helm T, Moll PJW, et~al. 2016.
\textit{Nature Physics} 12(10):916--919

\bibitem{jiang2023}
Jiang X, Qin M, Wei X, Xu L, Ke J, et~al. 2023.
\textit{Nature Physics} 19(3):365--371

\bibitem{nguyen2021}
Nguyen DH, Sidorenko A, Taupin M, Knebel G, Lapertot G, et~al. 2021.
\textit{Nature Communications} 12(1):4341

\bibitem{lee2023}
Lee K, Wang BY, Osada M, Goodge BH, Wang TC, et~al. 2023.
\textit{Nature} 619(7969):288--292

\bibitem{jaoui2022}
Jaoui A, Das I, {Di Battista} G, D{\'{i}}ez-M{\'{e}}rida J, Lu X, et~al. 2022.
\textit{Nature Physics} 18(6):633--638

\bibitem{lee2018}
Lee SS. 2018.
\textit{Annual Review of Condensed Matter Physics} 9(1):227--244

\bibitem{millis1993}
Millis AJ. 1993.
\textit{Phys. Rev. B} 48(10):7183--7196

\bibitem{altshuler1995}
Altshuler BL, Ioffe LB, Millis AJ. 1995.
\textit{Phys. Rev. B} 52(8):5563--5572

\bibitem{castellani1995}
Castellani C, Di~Castro C, Grilli M. 1995.
\textit{Phys. Rev. Lett.} 75(25):4650--4653

\bibitem{abanov2000}
Abanov A, Chubukov AV. 2000.
\textit{Phys. Rev. Lett.} 84(24):5608--5611

\bibitem{abanov2003}
Ar.~Abanov AVC, Schmalian J. 2003.
\textit{Advances in Physics} 52(3):119--218

\bibitem{pankov2004}
Pankov S, Florens S, Georges A, Kotliar G, Sachdev S. 2004.
\textit{Phys. Rev. B} 69(5):054426

\bibitem{chubukov2005}
Chubukov AV, Schmalian J. 2005.
\textit{Phys. Rev. B} 72(17):174520

\bibitem{lohneysen2007}
L\"ohneysen Hv, Rosch A, Vojta M, W\"olfle P. 2007.
\textit{Rev. Mod. Phys.} 79(3):1015--1075

\bibitem{metlitski2010b}
Metlitski MA, Sachdev S. 2010{\natexlab{a}}.
\textit{Phys. Rev. B} 82(7):075128

\bibitem{efetov2013}
Efetov KB, Meier H, P{\'{e}}pin C. 2013.
\textit{Nature Physics} 9(7):442--446

\bibitem{abrahams2014}
Abrahams E, Schmalian J, W\"olfle P. 2014.
\textit{Phys. Rev. B} 90(4):045105

\bibitem{meier2014}
Meier H, P\'epin C, Einenkel M, Efetov KB. 2014.
\textit{Phys. Rev. B} 89(19):195115

\bibitem{varma2015}
Varma CM. 2015.
\textit{Phys. Rev. Lett.} 115(18):186405

\bibitem{schlief2017}
Schlief A, Lunts P, Lee SS. 2017.
\textit{Phys. Rev. X} 7(2):021010

\bibitem{lunts2017}
Lunts P, Schlief A, Lee SS. 2017.
\textit{Phys. Rev. B} 95(24):245109

\bibitem{oganesyan2001}
Oganesyan V, Kivelson SA, Fradkin E. 2001.
\textit{Phys. Rev. B} 64(19):195109

\bibitem{metzner2003}
Metzner W, Rohe D, Andergassen S. 2003.
\textit{Phys. Rev. Lett.} 91(6):066402

\bibitem{lawler2006}
Lawler MJ, Barci DG, Fern\'andez V, Fradkin E, Oxman L. 2006.
\textit{Phys. Rev. B} 73(8):085101

\bibitem{rech2006}
Rech J, P\'epin C, Chubukov AV. 2006.
\textit{Phys. Rev. B} 74(19):195126

\bibitem{aji2007}
Aji V, Varma CM. 2007.
\textit{Phys. Rev. Lett.} 99(6):067003

\bibitem{zacharias2009}
Zacharias M, W\"olfle P, Garst M. 2009.
\textit{Phys. Rev. B} 80(16):165116

\bibitem{metlitski2010}
Metlitski MA, Sachdev S. 2010{\natexlab{b}}.
\textit{Phys. Rev. B} 82(7):075127

\bibitem{maslov2010}
Maslov DL, Chubukov AV. 2010.
\textit{Phys. Rev. B} 81(4):045110

\bibitem{dalidovich2013}
Dalidovich D, Lee SS. 2013.
\textit{Phys. Rev. B} 88(24):245106

\bibitem{fitzpatrick2014}
Fitzpatrick AL, Kachru S, Kaplan J, Raghu S. 2014.
\textit{Phys. Rev. B} 89(16):165114

\bibitem{lee1989gauge}
Lee PA. 1989.
\textit{Physical review letters} 63(6):680

\bibitem{halperin1993}
Halperin BI, Lee PA, Read N. 1993.
\textit{Phys. Rev. B} 47(12):7312--7343

\bibitem{polchinski1994low}
Polchinski J. 1994.
\textit{Nuclear Physics B} 422(3):617--633

\bibitem{nayak1994}
Nayak C, Wilczek F. 1994.
\textit{Nuclear Physics B} 430(3):534--562

\bibitem{chakravarty1995}
Chakravarty S, Norton RE, Sylju\aa{}sen OF. 1995.
\textit{Phys. Rev. Lett.} 74(8):1423--1426

\bibitem{bonesteel1996}
Bonesteel NE, McDonald IA, Nayak C. 1996.
\textit{Phys. Rev. Lett.} 77(14):3009--3012

\bibitem{lee2009}
Lee SS. 2009.
\textit{Phys. Rev. B} 80(16):165102

\bibitem{mross2010}
Mross DF, McGreevy J, Liu H, Senthil T. 2010.
\textit{Phys. Rev. B} 82(4):045121

\bibitem{holder2015}
Holder T, Metzner W. 2015.
\textit{Phys. Rev. B} 92(4):041112

\bibitem{balatsky1993}
Balatsky AV. 1993.
\textit{Philosophical Magazine Letters} 68(4):251--256

\bibitem{sudbo1995}
Sudb\o{} A. 1995.
\textit{Phys. Rev. Lett.} 74(13):2575--2578

\bibitem{yin1996}
Yin L, Chakravarty S. 1996.
\textit{International Journal of Modern Physics B} 10(07):805--845

\bibitem{son1999}
Son DT. 1999.
\textit{Phys. Rev. D} 59(9):094019

\bibitem{abanov2001}
Abanov A, Chubukov AV, Finkel'stein AM. 2001.
\textit{Europhysics Letters} 54(4):488

\bibitem{abanov1999}
Abanov A, Chubukov AV. 1999.
\textit{Phys. Rev. Lett.} 83(8):1652--1655

\bibitem{abanov2001b}
Abanov A, Chubukov AV, Schmalian J. 2001.
\textit{Europhysics Letters} 55(3):369

\bibitem{roussev2001}
Roussev R, Millis AJ. 2001.
\textit{Phys. Rev. B} 63(14):140504

\bibitem{abanov2004}
Abanov A, Chubukov A. 2004.
\textit{Phys. Rev. Lett.} 93(25):255702

\bibitem{she2009}
She JH, Zaanen J. 2009.
\textit{Phys. Rev. B} 80(18):184518

\bibitem{moon2010}
Moon EG, Chubukov A. 2010.
\textit{Journal of Low Temperature Physics} 161(1):263--281

\bibitem{levchenko2013}
Levchenko A, Vavilov MG, Khodas M, Chubukov AV. 2013.
\textit{Phys. Rev. Lett.} 110(17):177003

\bibitem{wang2013}
Wang Y, Chubukov A. 2013.
\textit{Phys. Rev. B} 88(2):024516

\bibitem{wang2015}
Wang Y, Chubukov AV. 2015.
\textit{Phys. Rev. B} 92(12):125108

\bibitem{varma2016}
Varma CM. 2016.
\textit{Reports on Progress in Physics} 79(8):082501

\bibitem{khodas2020}
Khodas M, Dzero M, Levchenko A. 2020.
\textit{Phys. Rev. B} 102(18):184505

\bibitem{lederer2015}
Lederer S, Schattner Y, Berg E, Kivelson SA. 2015.
\textit{Phys. Rev. Lett.} 114(9):097001

\bibitem{metlitski2015}
Metlitski MA, Mross DF, Sachdev S, Senthil T. 2015.
\textit{Phys. Rev. B} 91(11):115111

\bibitem{fitzpatrick2015}
Fitzpatrick AL, Kachru S, Kaplan J, Raghu S, Torroba G, Wang H. 2015.
\textit{Phys. Rev. B} 92(4):045118

\bibitem{raghu2015}
Raghu S, Torroba G, Wang H. 2015.
\textit{Phys. Rev. B} 92(20):205104

\bibitem{mandal2016}
Mandal I. 2016.
\textit{Phys. Rev. B} 94(11):115138

\bibitem{she2011}
She JH, Overbosch BJ, Sun YW, Liu Y, Schalm KE, et~al. 2011.
\textit{Phys. Rev. B} 84(14):144527

\bibitem{wang2017_eliash}
Wang H, Raghu S, Torroba G. 2017.
\textit{Phys. Rev. B} 95(16):165137

\bibitem{wang2018_thermal}
Wang H, Wang Y, Torroba G. 2018.
\textit{Phys. Rev. B} 97(5):054502

\bibitem{wang2016}
Wang Y, Abanov A, Altshuler BL, Yuzbashyan EA, Chubukov AV. 2016.
\textit{Phys. Rev. Lett.} 117(15):157001

\bibitem{wu2019}
Wu YM, Abanov A, Wang Y, Chubukov AV. 2019.
\textit{Phys. Rev. B} 99(14):144512

\bibitem{abanov2020-I}
Abanov A, Chubukov AV. 2020.
\textit{Phys. Rev. B} 102(2):024524

\bibitem{abanov2020-II}
Wu YM, Abanov A, Wang Y, Chubukov AV. 2020.
\textit{Phys. Rev. B} 102(2):024525

\bibitem{wu2020-III}
Wu YM, Abanov A, Chubukov AV. 2020.
\textit{Phys. Rev. B} 102(9):094516

\bibitem{wu2021-IV}
Wu YM, Zhang SS, Abanov A, Chubukov AV. 2021{\natexlab{a}}.
\textit{Phys. Rev. B} 103(2):024522

\bibitem{wu2021-V}
Wu YM, Zhang SS, Abanov A, Chubukov AV. 2021{\natexlab{b}}.
\textit{Phys. Rev. B} 103(18):184508

\bibitem{zhang2021-VI}
Zhang SS, Wu YM, Abanov A, Chubukov AV. 2021.
\textit{Phys. Rev. B} 104(14):144509

\bibitem{zhang2023}
Zhang SS, Chubukov AV. 2023.
\textit{Phys. Rev. Lett.} 131(8):086502

\bibitem{nosov2023}
Nosov PA, Burmistrov IS, Raghu S. 2023.
\textit{Phys. Rev. B} 107(14):144508

\bibitem{abanov2025}
Abanov A, Zhang SS, Chubukov AV. 2025.
\textit{Phys. Rev. B} 111(7):075157

\bibitem{berg2012}
Berg E, Metlitski MA, Sachdev S. 2012.
\textit{Science} 338(6114):1606--1609

\bibitem{schattner2016}
Schattner Y, Gerlach MH, Trebst S, Berg E. 2016{\natexlab{a}}.
\textit{Phys. Rev. Lett.} 117(9):097002

\bibitem{schattner2016_ising}
Schattner Y, Lederer S, Kivelson SA, Berg E. 2016{\natexlab{b}}.
\textit{Phys. Rev. X} 6(3):031028

\bibitem{dumitrescu2016}
Dumitrescu PT, Serbyn M, Scalettar RT, Vishwanath A. 2016.
\textit{Phys. Rev. B} 94(15):155127

\bibitem{gerlach2017}
Gerlach MH, Schattner Y, Berg E, Trebst S. 2017.
\textit{Phys. Rev. B} 95(3):035124

\bibitem{lederer2017}
Lederer S, Schattner Y, Berg E, Kivelson SA. 2017.
\textit{Proceedings of the National Academy of Sciences} 114(19):4905--4910

\bibitem{li2017}
Li ZX, Wang F, Yao H, Lee DH. 2017.
\textit{Phys. Rev. B} 95(21):214505

\bibitem{wang2017}
Wang X, Schattner Y, Berg E, Fernandes RM. 2017.
\textit{Phys. Rev. B} 95(17):174520

\bibitem{xu2017}
Xu XY, Sun K, Schattner Y, Berg E, Meng ZY. 2017.
\textit{Phys. Rev. X} 7(3):031058

\bibitem{wang2018}
Wang X, Wang Y, Schattner Y, Berg E, Fernandes RM. 2018.
\textit{Phys. Rev. Lett.} 120(24):247002

\bibitem{berg2019}
Berg E, Lederer S, Schattner Y, Trebst S. 2019.
\textit{Annual Review of Condensed Matter Physics} 10(1):63--84

\bibitem{klein2020}
Klein A, Chubukov AV, Schattner Y, Berg E. 2020.
\textit{Phys. Rev. X} 10(3):031053

\bibitem{xu2020}
Xu XY, Klein A, Sun K, Chubukov AV, Meng ZY. 2020.
\textit{npj Quantum Materials} 5(1):65

\bibitem{lunts2023}
Lunts P, Albergo MS, Lindsey M. 2023.
\textit{Nature Communications} 14(1):2547

\bibitem{patel2024_dqmc}
Patel AA, Lunts P, Albergo MS. 2024.
Strange metals and planckian transport in a gapless phase from spatially random
  interactions

\bibitem{esterlis2019}
Esterlis I, Schmalian J. 2019.
\textit{Phys. Rev. B} 100(11):115132

\bibitem{wang2020}
Wang Y. 2020.
\textit{Phys. Rev. Lett.} 124(1):017002

\bibitem{hauck2020}
Hauck D, Klug MJ, Esterlis I, Schmalian J. 2020.
\textit{Annals of Physics} 417:168120

\bibitem{wang2020b}
Wang Y, Chubukov AV. 2020.
\textit{Phys. Rev. Research} 2(3):033084

\bibitem{classen2021}
Classen L, Chubukov A. 2021.
\textit{Phys. Rev. B} 104(12):125120

\bibitem{esterlis2021}
Esterlis I, Guo H, Patel AA, Sachdev S. 2021.
\textit{Phys. Rev. B} 103(23):235129

\bibitem{guo2022}
Guo H, Patel AA, Esterlis I, Sachdev S. 2022.
\textit{Phys. Rev. B} 106(11):115151

\bibitem{patel2023}
Patel AA, Guo H, Esterlis I, Sachdev S. 2023.
\textit{Science} 381(6659):790--793

\bibitem{valentinis2023_lett}
Valentinis D, Inkof GA, Schmalian J. 2023{\natexlab{a}}.
\textit{Phys. Rev. B} 108(14):L140501

\bibitem{valentinis2023}
Valentinis D, Inkof GA, Schmalian J. 2023{\natexlab{b}}.
\textit{Phys. Rev. Res.} 5(4):043007

\bibitem{guo2024_cyclotron}
Guo H, Valentinis D, Schmalian J, Sachdev S, Patel AA. 2024.
\textit{Phys. Rev. B} 109(7):075162

\bibitem{li2024}
Li C, Valentinis D, Patel AA, Guo H, Schmalian J, et~al. 2024.
\textit{Phys. Rev. Lett.} 133(18):186502

\bibitem{pan2021}
Pan G, Wang W, Davis A, Wang Y, Meng ZY. 2021.
\textit{Phys. Rev. Res.} 3(1):013250

\bibitem{wang2021_dqmc}
Wang W, Davis A, Pan G, Wang Y, Meng ZY. 2021.
\textit{Phys. Rev. B} 103(19):195108

\bibitem{guo2024}
Guo H. 2024.
\textit{Phys. Rev. B} 110(15):155130

\bibitem{sutradhar2024}
Sutradhar J, Ruhman J, Klein A. 2024.
\textit{Phys. Rev. Res.} 6(4):L042036

\bibitem{sachdev1993}
Sachdev S, Ye J. 1993.
\textit{Phys. Rev. Lett.} 70(21):3339--3342

\bibitem{georges2000}
Georges A, Parcollet O, Sachdev S. 2000.
\textit{Phys. Rev. Lett.} 85(4):840--843

\bibitem{sachdev2010}
Sachdev S. 2010.
\textit{Phys. Rev. Lett.} 105(15):151602

\bibitem{kitaev2015}
Kitaev A. 2015{\natexlab{a}}.
\textit{Talks at KITP, University of California, Santa Barbara, Entanglement in
  Strongly-Correlated Quantum Matter}

\bibitem{kitaev2015b}
Kitaev A. 2015{\natexlab{b}}.
\textit{Talk 2}

\bibitem{sachdev2015}
Sachdev S. 2015.
\textit{Phys. Rev. X} 5(4):041025

\bibitem{maldacena2016}
Maldacena J, Stanford D. 2016.
\textit{Phys. Rev. D} 94(10):106002

\bibitem{polchinski2016}
Polchinski J, Rosenhaus V. 2016.
\textit{Journal of High Energy Physics} 2016(4):1

\bibitem{fu2017}
Fu W, Gaiotto D, Maldacena J, Sachdev S. 2017.
\textit{Phys. Rev. D} 95(2):026009

\bibitem{bi2017}
Bi Z, Jian CM, You YZ, Pawlak KA, Xu C. 2017.
\textit{Phys. Rev. B} 95(20):205105

\bibitem{song2017}
Song XY, Jian CM, Balents L. 2017.
\textit{Phys. Rev. Lett.} 119(21):216601

\bibitem{chowdhury2018}
Chowdhury D, Werman Y, Berg E, Senthil T. 2018.
\textit{Phys. Rev. X} 8(3):031024

\bibitem{chowdhury2022}
Chowdhury D, Georges A, Parcollet O, Sachdev S. 2022.
\textit{Rev. Mod. Phys.} 94(3):035004

\bibitem{sachdev2024}
Sachdev S. 2024.
\textit{International Journal of Modern Physics B} 38(32)

\bibitem{migdal1958}
Migdal A. 1958.
\textit{Sov. Phys. JETP} 7(6):996--1001

\bibitem{eliashberg1960}
Eliashberg G. 1960.
\textit{Sov. Phys. JETP} 11(3):696--702

\bibitem{marsiglio2020}
Marsiglio F. 2020.
\textit{Annals of Physics} 417:168102

\bibitem{Maldacena1998}
Maldacena JM. 1999.
\textit{Int. J. Theor. Phys.} 38:1113--1133

\bibitem{Witten1998}
Witten E. 1998.
\textit{Advances in Theoretical and Mathematical Physics} 2(2):253--291

\bibitem{Gubser1998}
Gubser S, Klebanov I, Polyakov A. 1998.
\textit{Physics Letters B} 428(1):105 -- 114

\bibitem{Inkof2021}
Inkof GA, Schalm K, Schmalian J. 2022.
\textit{npj Quantum Materials} 7(1)

\bibitem{2209.00474}
Schmalian J. 2022.
Holographic superconductivity of a critical fermi surface

\bibitem{hosseinabadi2023_thermalizationNFL}
Hosseinabadi H, Kelly SP, Schmalian J, Marino J. 2023.
\textit{Phys. Rev. B} 108(10):104319

\bibitem{kim2021dirac}
Kim J, Altman E, Cao X. 2021.
\textit{Physical Review B} 103(8):L081113

\bibitem{grunwald2024_light-inducedSC}
Grunwald L, Passetti G, Kennes DM. 2024.
\textit{Communications Physics} 7(1):79

\bibitem{cichutek2024_dissipativeSYK}
Cichutek N, R\"uckriegel A, Hansen MO, Kopietz P. 2024.
\textit{Phys. Rev. B} 109(15):155101

\bibitem{bashan2024_2-level-sys}
Bashan N, Tulipman E, Schmalian J, Berg E. 2024.
\textit{Phys. Rev. Lett.} 132(23):236501

\bibitem{tulipman2024solvable}
Tulipman E, Bashan N, Schmalian J, Berg E. 2024.
\textit{Physical Review B} 110(15):155118

\bibitem{Tikhanovskaya2022_maximalchaos}
Tikhanovskaya M, Sachdev S, Patel AA. 2022.
\textit{Phys. Rev. Lett.} 129(6):060601

\bibitem{davis2023_chaos}
Davis A, Wang Y. 2023.
\textit{Phys. Rev. B} 107(20):205122

\bibitem{wu2023_pdw}
Wu YM, Nosov PA, Patel AA, Raghu S. 2023.
\textit{Phys. Rev. Lett.} 130(2):026001

\bibitem{bashan2025extended}
Bashan N, Tulipman E, Kivelson SA, Schmalian J, Berg E. 2025.
\textit{arXiv preprint arXiv:2502.08699}

\bibitem{wang2023_density}
Wang X, Chowdhury D. 2023.
\textit{Phys. Rev. B} 107(12):125157

\bibitem{wang2024_landau-damp}
Wang X, Moessner R, Chowdhury D. 2024.
\textit{Phys. Rev. B} 109(12):L121102

\bibitem{patel2018}
Patel AA, Lawler MJ, Kim EA. 2018.
\textit{Phys. Rev. Lett.} 121(18):187001

\bibitem{gnezdilov2019}
Gnezdilov NV. 2019.
\textit{Phys. Rev. B} 99(2):024506

\bibitem{chowdhury2020}
Chowdhury D, Berg E. 2020{\natexlab{a}}.
\textit{Phys. Rev. Res.} 2(1):013301

\bibitem{wang2020_kuramoto}
Wang H, Chudnovskiy AL, Gorsky A, Kamenev A. 2020.
\textit{Phys. Rev. Res.} 2(3):033025

\bibitem{lantagne-hurtubise2021_spinfulSYK}
Lantagne-Hurtubise E, Pathak V, Sahoo S, Franz M. 2021.
\textit{Phys. Rev. B} 104(2):L020509

\bibitem{choi2022_SC}
Choi W, Tavakol O, Kim YB. 2022.
\textit{SciPost Phys.} 12:151

\bibitem{chudnovskiy2022_SC-insulator-transition}
Chudnovskiy AL, Kamenev A. 2022.
\textit{Phys. Rev. Lett.} 129(26):266601

\bibitem{gnezdilov2022_4e}
Gnezdilov NV, Wang Y. 2022.
\textit{Phys. Rev. B} 106(9):094508

\bibitem{li2023_SYKSC}
Li C, Sachdev S, Joshi DG. 2023.
\textit{Phys. Rev. Res.} 5(1):013045

\bibitem{abrikosov1960}
Abrikosov A, Gor'kov L. 1961.
\textit{Sov. Phys. JETP} 12:1243

\bibitem{georges2001}
Georges A, Parcollet O, Sachdev S. 2001.
\textit{Phys. Rev. B} 63(13):134406

\bibitem{schmalian1996}
Schmalian J, Langer M, Grabowski S, Bennemann K. 1996.
\textit{Computer Physics Communications} 93(2):141--151

\bibitem{andeson1959}
Anderson P. 1959.
\textit{Journal of Physics and Chemistry of Solids} 11(1):26--30

\bibitem{abrikosov1959}
Abrikosov A, Gorkov L. 1959.
\textit{Sov. Phys. JETP} 8(6):1090--1098

\bibitem{abrikosov1959b}
Abrikosov A, Gor’Kov L. 1959.
\textit{Sov. Phys. JETP} 9(1):220--221

\bibitem{potter2011}
Potter AC, Lee PA. 2011.
\textit{Phys. Rev. B} 83(18):184520

\bibitem{kang2016}
Kang J, Fernandes RM. 2016.
\textit{Phys. Rev. B} 93(22):224514

\bibitem{millis1988}
Millis AJ, Sachdev S, Varma CM. 1988.
\textit{Phys. Rev. B} 37(10):4975--4986

\bibitem{abanov2008}
Abanov A, Chubukov AV, Norman MR. 2008.
\textit{Phys. Rev. B} 78(22):220507

\bibitem{combescot1995}
Combescot R. 1995.
\textit{Phys. Rev. B} 51(17):11625--11634

\bibitem{kaplan2009}
Kaplan DB, Lee JW, Son DT, Stephanov MA. 2009.
\textit{Phys. Rev. D} 80(12):125005

\bibitem{esterlis_unpublished}
Esterlis I. unpublished

\bibitem{mahan2013}
Mahan GD. 2013.
Many-particle physics.
Springer Science \& Business Media

\bibitem{chubukov2020}
Chubukov AV, Abanov A, Esterlis I, Kivelson SA. 2020.
\textit{Annals of Physics} 417:168190Eliashberg theory at 60: Strong-coupling
  superconductivity and beyond

\bibitem{chowdhury2020_eliash}
Chowdhury D, Berg E. 2020{\natexlab{b}}.
\textit{Annals of Physics} 417:168125Eliashberg theory at 60: Strong-coupling
  superconductivity and beyond

\bibitem{aldape2022}
Aldape EE, Cookmeyer T, Patel AA, Altman E. 2022.
\textit{Phys. Rev. B} 105(23):235111

\bibitem{hertz1976}
Hertz JA. 1976.
\textit{Phys. Rev. B} 14(3):1165--1184

\bibitem{paul2017lattice}
Paul I, Garst M. 2017.
\textit{Physical Review Letters} 118(22):227601

\bibitem{michon2023}
Michon B, Berthod C, Rischau CW, Ataei A, Chen L, et~al. 2023.
\textit{Nature Communications} 14(1):3033

\bibitem{patel2024_localized-bosons}
Patel AA, Lunts P, Sachdev S. 2024.
\textit{Proceedings of the National Academy of Sciences} 121(14):e2402052121

\bibitem{dellanna2006}
Dell'Anna L, Metzner W. 2006.
\textit{Phys. Rev. B} 73(4):045127

\bibitem{damia2020}
Damia JA, Sol\'{\i}s M, Torroba G. 2020.
\textit{Phys. Rev. B} 102(4):045147

\bibitem{damia2021}
Damia JA, Sol\'{\i}s M, Torroba G. 2021.
\textit{Phys. Rev. B} 103(15):155161

\bibitem{varma1989}
Varma CM, Littlewood PB, Schmitt-Rink S, Abrahams E, Ruckenstein AE. 1989.
\textit{Phys. Rev. Lett.} 63(18):1996--1999

\bibitem{khurana1990}
Khurana A. 1990.
\textit{Phys. Rev. Lett.} 64(16):1990--1990

\bibitem{Sachdev2000}
Sachdev S. 2000.
{Quantum Phase Transitions}.
Cambridge: Cambridge University Press

\bibitem{varma2020}
Varma CM. 2020.
\textit{Rev. Mod. Phys.} 92(3):031001

\bibitem{mitrano2018}
Mitrano M, Husain AA, Vig S, Kogar A, Rak MS, et~al. 2018.
\textit{Proceedings of the National Academy of Sciences} 115(21):5392--5396

\bibitem{husain2019}
Husain AA, Mitrano M, Rak MS, Rubeck S, Uchoa B, et~al. 2019.
\textit{Phys. Rev. X} 9(4):041062

\bibitem{klein_toappear}
Klein A, Schmalian J. forthcoming

\bibitem{DeHaro2001}
{De Haro} S, Skenderis K, Solodukhin SN. 2001.
\textit{Communications in Mathematical Physics} 217(3):595--622

\bibitem{Skenderis_2002}
Skenderis K. 2002.
\textit{Classical and Quantum Gravity} 19(22):5849--5876

\bibitem{Donos2014}
Donos A, Gauntlett JP. 2014.
\textit{Journal of High Energy Physics} 2014(11):81

\bibitem{PhysRevB.76.144502}
Hartnoll SA, Kovtun PK, M\"uller M, Sachdev S. 2007.
\textit{Phys. Rev. B} 76(14):144502

\bibitem{casalderrey-solana_liu_mateos_rajagopal_wiedemann_2014}
Casalderrey-Solana J, Liu H, Mateos D, Rajagopal K, Wiedemann UA. 2014.
Gauge/string duality, hot qcd and heavy ion collisions.
Cambridge University Press

\bibitem{PhysRevLett.94.111601}
Kovtun PK, Son DT, Starinets AO. 2005.
\textit{Physical Review Letters} 94(11):23--26

\bibitem{PhysRevLett.87.081601}
Policastro G, Son DT, Starinets AO. 2001.
\textit{Phys. Rev. Lett.} 87(8):081601

\bibitem{Gubser2008}
Gubser SS. 2008.
\textit{Physical Review D - Particles, Fields, Gravitation and Cosmology} 78(6)

\bibitem{Hartnoll2008}
Hartnoll SA, Herzog CP, Horowitz GT. 2008{\natexlab{a}} (Dc)

\bibitem{Hartnoll_2008HEP}
Hartnoll SA, Herzog CP, Horowitz GT. 2008{\natexlab{b}}.
\textit{Journal of High Energy Physics} 2008(12):015--015

\bibitem{Donos2011}
Donos A, Gauntlett JP. 2011.
\textit{Journal of High Energy Physics} 2011(8)

\bibitem{PhysRevB.96.195128}
Delacr\'etaz LV, Gout\'eraux B, Hartnoll SA, Karlsson A. 2017.
\textit{Phys. Rev. B} 96(19):195128

\bibitem{1603.03029}
Amoretti A, Are{\'{a}}n D, Gout{\'{e}}raux B, Musso D. 2018.
\textit{Physical Review D} 97(8):1--18

\bibitem{Breitenlohner1982}
Breitenlohner P, Freedman DZ. 1982.
\textit{Physics Letters B} 115(3):197--201

\bibitem{inkof2022quantum}
Inkof GA. 2022.
\textit{PhD-thesis, Karlsruhe Institute of Technology}

\bibitem{Das2018}
Das SR, Ghosh A, Jevicki A, Suzuki K. 2018.
\textit{JHEP} 07:184

\bibitem{Sachdev2019}
Sachdev S. 2019.
\textit{Journal of Mathematical Physics} 60(5):0--23

\bibitem{Luttinger1961}
Luttinger JM. 1960.
\textit{Phys. Rev.} 119(4):1153--1163

\bibitem{Gu2020}
Gu Y, Kitaev A, Sachdev S, Tarnopolsky G. 2020.
\textit{Journal of High Energy Physics} 2020(2):157

\bibitem{Faulkner2011}
Faulkner T, Iqbal N, Liu H, McGreevy J, Vegh D. 2011

\bibitem{gorkov1959}
Gor’kov LP. 1959.
\textit{Sov. Phys. JETP} 9(6):1364--1367

\bibitem{ginzburg2009theory}
Ginzburg VL, Ginzburg VL, Landau L. 2009.
On the theory of superconductivity.
Springer

\bibitem{zhang2024}
Zhang SS, Raines ZM, Chubukov AV. 2024.
\textit{Phys. Rev. B} 109(24):245132

\bibitem{hardy2025}
Hardy A, Parcollet O, Georges A, Patel AA. 2025.
\textit{Physical Review Letters} 134(3)

\end{thebibliography}

\end{document}